\documentclass[reprint,superscriptaddress, amsmath,amssymb,aps,prl,floatfix]{revtex4-2}

\usepackage{color}
\usepackage[dvipsnames]{xcolor}

\usepackage{graphicx}
\usepackage{dcolumn}
\usepackage{bm}

\begin{document}

\preprint{APS/123-QED}

\title{Magneto-Thermoelectric Transport in\\ Graphene Quantum Dot with Strong Correlations}

\author{Laurel E. Anderson}
\affiliation{Department of Physics, Harvard University, Cambridge, Massachusetts 02138, USA}
\author{Antti Laitinen}
\affiliation{Department of Physics, Harvard University, Cambridge, Massachusetts 02138, USA}
\author{Andrew Zimmerman}
\affiliation{Department of Physics, Harvard University, Cambridge, Massachusetts 02138, USA}
\author{Thomas Werkmeister}
\affiliation{Department of Applied Physics, Harvard University, Cambridge, Massachusetts 02138, USA}
\author{Leyna Shackleton}
\affiliation{Department of Physics, Harvard University, Cambridge, Massachusetts 02138, USA}
\author{Alexander Kruchkov}
\affiliation{Department of Physics, Harvard University, Cambridge, Massachusetts 02138, USA}
\affiliation{Department of Physics, Princeton University, Princeton, New Jersey 08544, USA}
\affiliation{Institute of Physics, \'Ecole Polytechnique F\'ed\'erale de Lausanne, Lausanne, CH 1015, Switzerland; and Branco Weiss Society in Science, ETH Zurich, Zurich, CH 8092, Switzerland}
\author{Takashi Taniguchi}
\affiliation{Research Center for Materials Nanoarchitectonics, National Institute for Materials Science,  1-1 Namiki, Tsukuba 305-0044, Japan}
\author{Kenji Watanabe}
\affiliation{Research Center for Electronic and Optical Materials, National Institute for Materials Science, 1-1 Namiki, Tsukuba 305-0044, Japan}
\author{Subir Sachdev}
\affiliation{Department of Physics, Harvard University, Cambridge, Massachusetts 02138, USA}
\author{Philip Kim}
\affiliation{Department of Physics, Harvard University, Cambridge, Massachusetts 02138, USA}
\affiliation{Department of Applied Physics, Harvard University, Cambridge, Massachusetts 02138, USA}
 \email{pkim@physics.harvard.edu}


\begin{abstract}
Disorder at the etched edges of graphene quantum dots (GQD) enables random all-to-all interactions between localized charges in partially-filled Landau levels, providing a potential platform to realize the Sachdev-Ye-Kitaev (SYK) model. We use quantum Hall edge states in the graphene electrodes to measure electrical conductance and thermoelectric power across the GQD. We observe a rapid diminishing of electric conductance fluctuations and slowly decreasing thermoelectric power across the GQD with increasing temperature, consistent with recent theoretical predictions for the SYK regime.

\end{abstract}

\maketitle

Strong electronic correlations can generate an emergent system that hosts collective excitations without quasiparticles, deviating from the conventional Fermi liquid picture. One proposed description is the Sachdev-Ye-Kitaev (SYK) model, characterized by random, all-to-all four-body interactions. Originally a model for strange metals and complex quantum phases~\cite{Sachdev1993}, this model also has been shown to be holographically dual to theories of quantum gravity~\cite{Kitaev2015,Sachdev2010,Chowdhury2022}, prompting searches for an experimental, solid-state realization of the SYK model~\cite{Franz2018}.

Generating an SYK state requires many electrons at the same energy with random all-to-all interactions. A theoretical proposal suggests creating these conditions by applying an external magnetic field to a graphene quantum dot (GQD) with an irregular boundary \cite{Chen2018,Brzezinska2022}. 
The dispersionless nature of Landau levels (LLs) on the lattice allow the electrons inside the GQD to remain nearly degenerate, despite the presence of edge disorder.
The irregular shape of the GQD edge causes the electronic wavefunctions to acquire a random spatial structure, creating random all-to-all interactions between the degenerate fermions in the dot, precisely as needed for the SYK model. 

Experimentally, it has been shown that the charge transport across etch-defined GQDs often exhibits the emergence of chaotic dynamics, as a result of the combination of confinement and disorder \cite{Ponomarenko2008,Engels2013}. Detailed theoretical modeling \cite{Brzezinska2022} suggests that an etch-defined, nanoscale GQD subjected to quantizing out-of-plane magnetic fields of 10-20~T may host strongly-correlated dynamics reminiscent of the SYK model. 
Due to the non-Fermi liquid (NFL) nature of the SYK system, transport through SYK GQDs can produce distinctive characteristic behavior compared to a Fermi liquid (FL) description. For example, nonvanishing extensive entropy in the low-temperature in a SYK QD produces temperature-independent, non-vanishing thermoelectric power (TEP), strongly deviating from the conventional Mott prediction in the FL regime~\cite{Kruchkov2020}.
Electrical conductance fluctuations, which in the FL regime are large and governed by single-particle random matrix theory, are suppressed in the SYK regime, a result of the absence of quasiparticle excitations~\cite{Shackleton2023}.
Since FL-to-NFL transition in the GQD can be tuned by magnetic field and temperature~\cite{Can2018}, temperature- and field-dependent transport through the dot can be utilized to investigate emergence of SYK physics in this system.   

In this work, we study the interplay of disorder, spatial confinement and strong electronic interactions in disordered GQD subjected to quantizing magnetic fields of up to 10~T. We measure electrical conductance and TEP across the GQD as a function of temperature, identifying a low-temperature FL phase and high-temperature NFL phase separated by a transition regime. We observe strong suppression of electrical conductance fluctuations and nearly temperature-independent TEP in the NFL regime, consistent with theoretical expectations for the SYK model.

\begin{figure}
    \includegraphics[width=0.45\textwidth]{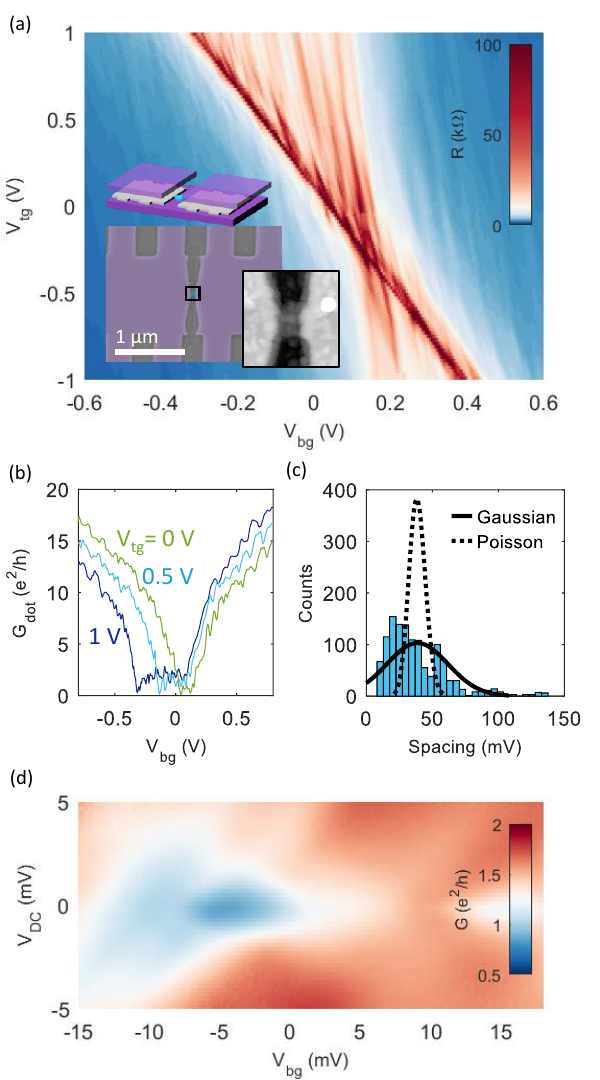}
    \caption{(a): $R_{dot}$ as a function of $V_{bg}$ and $V_{tg}$ at zero applied magntic fields, $T_{bath}$ = 350 mK, and $V_{Si}$ = 28 V. Inset, above: schematic of dot device with continuous bottom graphite gate (purple), GQD (blue) connected to larger reservoirs (gray), and separate top graphite gates above each reservoir. Below: scanning electron microscope image of dot device and atomic force microscope image of the GQD region. (b) $G_{\rm dot}$ as a function of $V_{bg}$ at $V_{tg}= 1$~V (dark blue), 0.5 V (light blue), and 0 V (green). (c) histogram of spacing between $G_{\rm dot}$ minima, with fits of Poisson distribution (dashed line) and Gaussian distribution (dashed line). (d) $G_{dot}$ as a function of $V_{bg}$ and DC bias $V_{DC}$ at $B=0$~T and $V_{tg} = 0.5$~V.}
    \label{fig1}
\end{figure}

The inset of Figure \ref{fig1}(a) shows a schematic diagram and electron microscope picture of a GQD used in this study. The device consists of hBN-encapsulated monolayer graphene with top and bottom graphite gates, fabricated using standard polymer stacking techniques \cite{Dean2010,Wang2013}. We shape the heterostructure into a Hall bar geometry using reactive ion etching, then etch a constriction with a $\sim$100 nm diameter island in the center, dividing the active region of the device into two large reservoirs that act as external contacts coupled to the central graphene dot. The top graphite gate above the constriction has been removed to enable independent tuning of the charge carrier densities in the GQD and graphene reservoirs. We note that the bottom graphite is separated from GQD with a thin (5.1 nm) hBN layer in order to reduce the Coulomb charging energy. 

The resistance measured across the dot ($R_{dot}$) is measured by biasing the GQD using the graphene reservoir electrodes. Figure \ref{fig1}(a) shows $R_{dot}$ as a function of the bottom and top graphite gate voltages, $V_{bg}$ and $V_{tg}$. The main diagonal feature in this plot corresponds to the charge neutrality point (CNP) of the graphene reservoirs. Near this reservoir CNP line, an array of steeper vertical features strongly controlled by $V_{bg}$ arise from conductance fluctuations in the dot. Due to our device structure, it is expected that the GQD is more strongly coupled to the bottom gate than the top gate. Line cuts of the dot conductance in Figure \ref{fig1}(b) highlight these features, which resemble previous studies of etched GQDs \cite{Ponomarenko2008,Guttinger2009}. We also observe that the $R_{dot}$ is maximized ($\sim$100~k$\Omega$) when both the reservoirs and the GQD are at their respective CNPs. We can identify 4 different segmented regions bounded by the CNP lines of the reservoirs and GQD in the $V_{bg}$--$V_{tg}$ plane. Generally we find $R_{dot}$ is larger in the $npn$ (reservoirs $n$-doped and GQD $p$-doped) or $pnp$ regimes compared to $nnn$ and $ppp$ regimes due to the reduced coupling of GQD to the reservoirs when their charge carrier polarities are opposite. 

A histogram of conductance minima spacing in the $npn$ regime (Fig. \ref{fig1}(c)) shows better resemblance to Gaussian than Poissonian statistics, suggesting chaotic dynamics~\cite{Ponomarenko2008,Huang2011}. Figure \ref{fig1}(d) shows a bias voltage ($V_{DC}$) and gate voltage  dependent conductance map in a suppressed conductance region of  Fig.~\ref{fig1}(b), where the GQD is near its CNP while the reservoirs are $n$-doped. We find the conductance in this stability diagram remains finite and lacks sharp Coulomb blockade features, suggesting the charging energy of the GQD is much smaller than the experimental base temperature of 350~mK. This is consistent with our design parameters for the device.

\begin{figure}
    \includegraphics[width=0.44\textwidth]{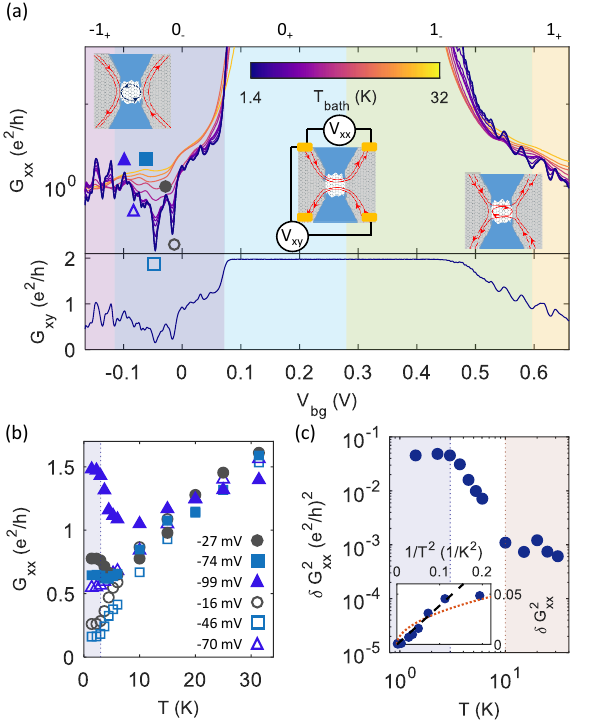}
    \caption{(a) Upper panel: $G_{xx}$ at $B=10$~T with $V_{bg}$ and $V_{tg}$ simultaneously varied to maintain $\nu_{\rm res}$= 2, at a range of temperatures between 1.4 K and 32 K as indicated by the color scale. Shaded regions show the doping regions for various Landau levels in the dot. Inset schematics illustrate the general behavior of the edge states in different doping regions, as well as the voltages measured to determine $G_{xy}$ and $G_{xx}$. Open symbols mark minima plotted in (b). Lower panel: $G_{xy}$ measured along the same $V_{bg}$ and $V_{tg}$ values as $G_{xx}$ at $T=1.41$~K. (b) The three lowest minima (open symbols) and nearby peaks (filled symbols) of $G_{xx}$ in the n=0 Landau level at $B = 10$~T as a function of temperature. Blue dashed line marks onset of Fermi liquid behavior. (c) Variance of $G_{xx}$ in n$_-$ Landau level at $B=10$~T. Blue dashed line marks $T_1$, while orange dashed line marks $T_2$. Inset: Variance of $G_{xx}$ versus $1/T^2$. Black dashed line and orange dotted line show $1/T^2$ and $1/T$ fits, respectively, for $T>3$~K.}
    \label{fig2}
\end{figure}

Upon applying a strong perpendicular magnetic field, $B$, the wide graphene reservoir regions show a robust onset of the quantum Hall (QH) effect (see Supplemental Material (SM)). We use the QH edge states in the reservoir regions to deliver charge current $I$ to the GQD. As shown in the insets of Figure \ref{fig2}(a), we measure the longitudinal (transverse) voltage $V_{xx}$ ($V_{xy}$). The upper and lower panels of Figure \ref{fig2}(a) show the corresponding longitudinal (transverse) conductance $G_{xx}=I/V_{xx}$ ($G_{xy}=I/V_{xy}$) as a function of $V_{bg}$. Here, we keep the graphene reservoirs at constant filling fraction $\nu=2$ by adjusting the top and bottom gate voltages simultaneously. In these measurements, we observe three principal zones of behavior: (1) suppressed conductance when the dot and reservoirs have opposite carrier types (i.e., $npn$ or $pnp$ regimes); (2) full transmission of integer QH edge states (i.e., the GQD is in $\nu_{dot}=2$ QHE regime, resulting in $G_{xy}=2e^2/h$ in Fig. \ref{fig2}(a)); and (3) reentrance of finite conductance fluctuations where $\nu_{dot}>2$. Combining this gate-dependent transport data with the TEP measured across the GQD (see Fig. \ref{fig4}(a), which will be discussed later), we can identify the gate voltage regions corresponding to the Landau level (LL) specified by $n_\pm$, where $n$ is the LL index and subscript $+(-)$ corresponds to the electron (hole) side of the LL. 

Since the LL filling fraction of the graphene reservoir regions is kept at $\nu_{res}=2$, the GQD is weakly coupled to graphene electrodes for $\nu_{dot}<\nu_{res}$. This condition prevents the highly-conductive QH edge states from shorting the graphene reservoirs, allowing us to study charge transport through the GQD. The QH edge states in the reservoir serve as few-mode FL electrodes, tunnel coupled to the GQD. Employing a small number of FL modes to probe the GQD is important for preserving signatures of SYK physics, as coupling an SYK dot to a large number of FL modes is predicted to disrupt the SYK phase~\cite{Can2018,Banerjee2017}. In this transport regime, where the GQD filling changes from $-1_+$ to $0_-$, we find $G_{xx}$ exhibits large fluctuations as $V_{bg}$ is changed. As the temperature increases, these fluctuations diminish toward a smoothly and slowly-varying background value, as shown in the upper panel of Figure \ref{fig2}(a). To highlight the temperature-dependent electrical conductance changes in the SYK transport regime, Figure \ref{fig2}(b) shows the temperature dependence of local extrema of $G_{xx}(V_{bg})$ in the GQD $0_-$ regime, with specific minima (maxima) marked by open (closed) symbols in Figure \ref{fig2}(a). We find that the temperature dependence of the local minima of $G_{xx}$ is nearly flat for temperature $T<$ 3~K, then linearly increasing at higher temperatures. Local conductance maxima in the same transport regime similarly show nearly-constant magnitudes up to $\sim3$~K, drop toward the values of the minima as temperature increases to $\sim10$~K, and increase approximately linearly as the temperature increases further. 

To quantify the temperature dependence GQD conductance fluctuation, we study the variance of the conductance $\delta G_{xx}^2$ within transport regime $0_-$ after subtracting the broadly-modulated baseline value. Figure \ref{fig2}(c) shows $\delta G_{xx}^2$ in the temperature range between 1.4 K and 30 K. This analysis highlights two relevant transition temperatures identified in the behavior of $G_{xx}(T)$ discussed above: while in the low temperature limit $T<T_1 \approx 3$~K, $\delta G_{xx}^2(T)$ is nearly constant, for $T_1<T<T_2\approx10$ K, $\delta G_{xx}^2$ decreases rapidly, then less steeply for $T>T_2$.

Recent theoretical work\cite{Brzezinska2022,Shackleton2023} has predicted strong suppression of $\delta G_{xx}^2(T)$ in SYK QDs coupled to FL reservoirs. In the presence of single-particle hopping energy $t$ between the localized states, SYK physics can be realized when the temperature is smaller than the coherence energy $E_{coh} = t^2/J$, where $J$ is the strength of all-to-all interactions in the SYK dot. Here, the theory predicts $\delta G_{xx}^2 \sim T^{-1}$ for $k_B T \ll E_{coh}$, crossing over to $\delta G_{xx}^2\sim T^{-2}$ for $k_B T \gg E_{coh}$~\cite{Shackleton2023}. As shown in Figure \ref{fig2}(c), the experimentally-observed variance exhibits $\delta G_{xx}^2\sim T^{-2}$ in the high-temperature limit (blue dashed line in inset), followed by $\delta G_{xx}^2\sim T^{-2}$ in the intermediate temperature regime, before saturating in the low temperature limit. The strong suppression ($\sim T^{-2}$) of the conductance fluctuations that we observe is a potential hallmark of SYK dynamics in the GQD, although the exact predicted temperature dependence is contingent on the coupling between the GQD and the reservoirs~\cite{Can2018,Shackleton2023}.

\begin{figure}
    \includegraphics[width=0.45\textwidth]{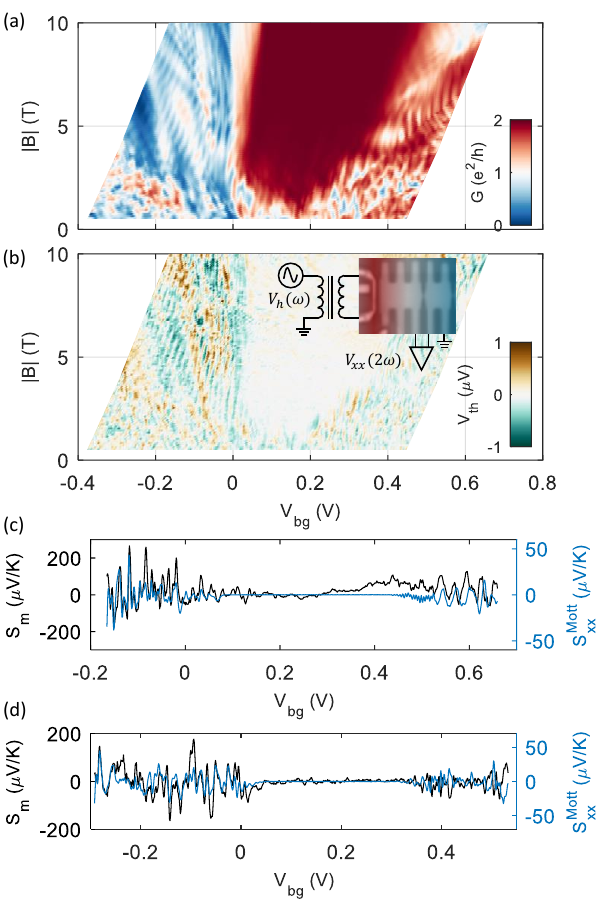}
    \caption{(a) Evolution of $G_{xy}$ at $T_{bath}=3$ K as a function of $V_{bg}$ and $B$, with $V_{tg}$ simultaneously varied to keep the reservoirs at constant filling $\nu_{res}$=2. (b) Evolution of thermally-induced voltage $V_{th}$ with the same experimental parameters as (a), using constant heater voltage $V_{h}$=0.3 V. Inset: schematic of TEP measurement circuit overlaid on optical image of device. (c) Comparison line scans of $S_m$ (black, left y-axis) and Mott formula calculation (blue, right y-axis) along $\nu_{res}$=2 line at $B=10$ T, $T=3$ K. (d) Same comparison as (c), but at $B=4$ T, with corresponding shift of the gate voltage values to maintain $\nu_{res}$=2.}
    \label{fig3}
\end{figure}

\begin{figure}
    \includegraphics[width=0.45\textwidth]{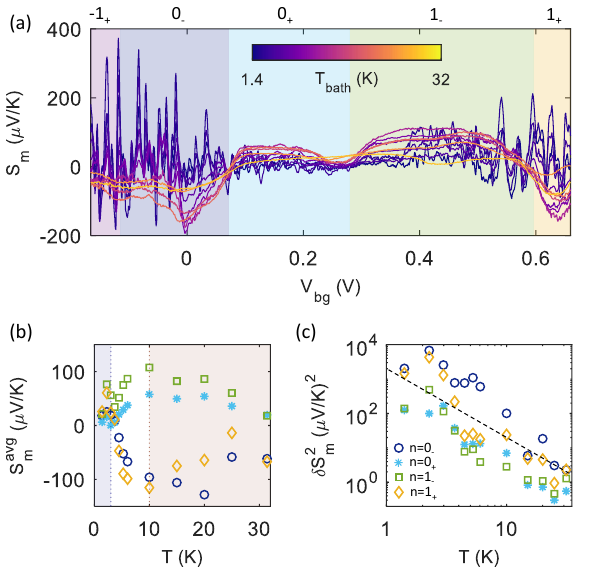}
    \caption{(a): TEP $S_{m}$ at $B=10$ T with $\nu_{res}$=2 at a range of temperatures between 1.4 K and 32 K. Shading indicates doping regions for various Landau levels in the GQD as defined in Fig.~2. (b): Average value of TEP $S^{avg}_m$ as a function of bath temperature $T$ in the regions highlighted by colors in (a). (c) TEP variance of $\delta S_{m}^2$ as a function of bath temperature for the regions highlighted in (a). Dashed line shows $1/T^2$ scaling.}
    \label{fig4}
\end{figure}

The strong suppression of conductance fluctuations in the GQD described above spurs us to investigate its thermoelectric response in similar transport regimes, in search of a more distinctive signature of the emergence of SYK physics~\cite{Kruchkov2020,Shackleton2023,Davison2017}. Here, we apply an AC bias $V_{h}(\omega)$ at frequency $\omega$ to a substrate heater at the edge of one of the graphene reservoirs (inset of Fig. \ref{fig3}(b)). The heating current generates a temperature gradient across the device modulated at frequency $2\omega$. By measuring the voltage response across the GQD at frequency $2\omega$, we obtain the thermoelectric voltage $V_{th}=\sqrt{2}V_{xx}(2\omega)$ in response to the temperature difference $\Delta T$ across the GQD. Figure \ref{fig3}(a-b) provides a comparison of the magnetic field dependence of $G_{xy}$ (measured in the center of the $\nu_{\rm res}=2$ plateau at each field) with the thermally-induced voltage $V_{th}$. In the magnetic field-dependent $G_{xy}$ measurement (Fig. \ref{fig3}(a)), both reservoir edge states are transmitted through the GQD in a wide range of densities down to $|B|\sim$ 3 T. The $npn$ and $nnn$ regimes show shifting patterns of oscillations as a function of $B$ and the carrier density in the GQD, reminiscent of previous studies of larger quantum Hall $pn$ and $npn$ junctions\cite{Wei2018,Williams2007,Abanin2007,Ozyilmaz2007,Wang2019}. At lower magnetic fields, the region of maximal conductance through the dot shrinks and the transport becomes completely dominated by fluctuations. The thermally-induced voltage $V_{th}$ measured under the same conditions (Fig. \ref{fig3}(b)) exhibits many of similar features, suggesting a strong correlation between the conductance and TEP in the GQD.
 
To obtain the TEP of the dot, $S_m=-V_{th}/\Delta T$, we need to estimate $\Delta T$ across the GQD for a given heater bias $V_h$. We employ temperature-dependent $R_{xx}$ minima in the QHE regime, measured at a pair of contacts in the graphene reservoirs. The minima of $R_{xx}$ are lifted as a function of the thermal bath temperature $T_{bath}$ and $V_{h}$, which allow us to estimate the temperature difference $\Delta T$ across the GQD, after considering the device geometry (see SM for details).  

As a benchmark, we compare the measured TEP with a generalized version of the conventional Mott formula \cite{Zuev2008},
\begin{equation}\label{eq:genMott}
    S^{Mott}_{ij}=-\frac{\pi^2}{3}\frac{k_B^2}{e} T G_{il}^{-1} \left[ \frac{\partial G_{lj}}{\partial \mu}\right],
\end{equation}
where $G_{ij}$ is the electrical conductance tensor and $\mu$ is the chemical potential.
At $B=10$~T (Fig. \ref{fig3}(c)), these quantities are broadly similar, but their magnitudes differ significantly. There is greater qualitative resemblance at lower magnetic field $B=4$~T (Fig. \ref{fig3}(d)), but the magnitudes of $S_m$ is larger than $S^{Mott}_{xx}$. 

An important contributing factor to the magnitude discrepancy is likely that $\Delta T$ is calculated based on measurements of the nearest pair of voltage leads in the two reservoirs, which is necessarily larger than the temperature gradient across the dot itself. In contrast, $V_{th}$ arises almost entirely in the dot, since the reservoirs are kept at $\nu_{res}$ = 2, which does not contribute to the TEP in this experimental configuration~\cite{Checkelsky2009,Zuev2008,Kermani2014,Duan2018}. As such, it is to be expected that $S_m$ overestimates the true TEP of the GQD. 
For these reasons, our observation $|S_{m}|>|S^{Mott}_{xx}|$ is consistent with expectations. In the following analysis, we discuss trends in the TEP that are not impacted by questions of geometric rescaling.

Examining the temperature-dependent TEP reveals many of the similar relevant energy scales and regimes of behavior as we observe in the electrical conductance. Figure \ref{fig4}(a) shows $S_m(T)$ as a function of $V_{bg}$ at $B=10$~T, at temperatures between 1.4 and 32~K. In the low temperature limit, $S_m$ follows the same gate dependence trend as $S_{xx}^{Mott}$, as discussed above. Particularly, $S_m \approx 0$ in the region 0.1$< V_{bg} <$0.4 V. This gate range is where we observe $G_{xy}=2e^2/h$, indicating both the GQD and graphene reservoirs are in the $\nu=2$ QH state. Vanishing TEP in the regime of the QH plateau is in agreement with previous studies~\cite{Zuev2008,Wei2009,Checkelsky2009,Kermani2014,Ghahari2016}. Outside of this QHE plateau region, however, $S_m$ and $S_{xx}^{Mott}$ exhibit rapid oscillations. The fact that the pattern of fluctuations can be explained by Eq.~\ref{eq:genMott} suggests that the resonance transport across the GQD is responsible for these rapid changes of $S_m$ as a function of $V_{bg}$. 

As $T$ increases, similar to the higher-temperature behavior of $G_{ij}$, the fluctuations of $S_m$ as a function of gate voltage are suppressed. In the high-temperature regime ($T \gtrsim 10$~K), $S_m \approx 0$ at half-filling of the LLs (i.e, gate voltage corresponding to the transitions between $0_-$ and $0_+$ for $n=0$ LL and between $1_-$ and $1_+$ for $n=1$ LL), due to the particle-hole symmetry across the LL in the GQD. We also observe $S_m \approx 0$ at the transition between $0_+$ and $1_-$ for $T< 30$~K, which corresponds to the center of the well-developed QHE plateau at $\nu=2$ across the entire device. Away from these vanishing points of TEP, $S_m$ varies smoothly. We found, however, unlike the low-temperature regime ($T \lesssim 3$~K), $S_m(V_{bg})$ does not follow the trend of $S^{Mott}_{xx}(V_{bg})$ at high temperatures ($T \gtrsim 10$~K), suggesting the breakdown of the single-particle picture described by the Mott formula (see SM for additional data).

Fig.~\ref{fig4}(b) shows $S^{avg}_m(T)$, the temperature-dependent averaged $S_m$ within the gate voltage regions corresponding to $n_{\pm}$. In all gate regimes, we find that $|S^{avg}_m(T)|$ exhibits distinctively different characteristic temperature-dependent behaviors. For $T<T_1$, $|S^{avg}_m(T)|$ changes relatively slowly as large gate-dependent fluctuations dominate, similar to the low-temperature regime of $G_{xx}(T)$ discussed previously. For $T_1<T<T_2$, $|S^{avg}_m(T)|$ monotonically increases as $T$ increases until it reaches the maximum value near $T_2$, and then decreases slowly as temperature increases further. We further compute the variance of the TEP, $\delta S^2$, in each sector (Fig. \ref{fig4}(c)). We find that $\delta S^2$ decreases as $T$ increases, generally consistent with $1/T^2$ scaling across the entire measured range.

The complete breakdown of the Mott formalism in the high-temperature regime, particularly the slowly decreasing $|S^{avg}_m(T)|$ we observe for $T>T_2$, is inconsistent with conventional FL physics in the GQD. Indeed, recent numerical modeling of the GQD thermoelectric properties~ \cite{Shackleton2023} predicts that the TEP in the low-temperature FL regime transitions into the SYK regime at higher temperatures, where slowly decreasing TEP is expected. Our experimental observations bear a resemblance to these theoretical predictions, suggesting an identification of $k_BT_1$ as related to the energy scale of coupling to the reservoirs and $k_BT_2 \sim E_{coh}$. However, we note that the temperature dependence of $\delta S^2$ is inconsistent with the predictions of $1/T$ scaling in this theoretical study. Some of these discrepancies from the theory, particularly in SYK regime, may be related to the relatively small population of SYK modes in the GQD~\cite{Kruchkov2020}. The approximate number of localized states at $B=10$ T is $N=BA_{dot}/\Phi_0\approx 33$, where $A_{dot}$ is the area of the GQD. While $N\gg 1$, it is still far from the conformal limit ($N\rightarrow \infty$), necessitating the inclusion of higher-order terms to fully account for the temperature scaling behavior, and the consideration of $J/N$ as another relevant energy scale. It is also possible that the strength of $J$, which we cannot directly measure, is significantly smaller than recent theoretical assumptions \cite{Can2018,Kruchkov2020,Brzezinska2022}. In this case, the system may never access the conformal limit of the SYK model, instead inhabiting a crossover regime between Fermi liquid and SYK dynamics \cite{Altland2019}. 

In conclusion, we have fabricated GQDs with suppressed single-electron charging energy. Under strong magnetic fields, edge disorder alters charge transport in the strongly-correlated electronic system at elevated temperatures. We observe temperature-dependent conductance fluctuation and thermoelectric power that exhibit transitioning behaviors from the FL to the putative SYK regime. Further experimental and theoretical studies, particularly considering the effects of varied coupling between the FL leads and the GQD, may distinguish between the emergence of an SYK phase and alternative scenarios, such as disordered $pn$ junction network formation~\cite{Wei2018,Wang2019} in the disordered GQD under magnetic fields. 
For more comprehensive statistics on disorder averaging, it will be beneficial to undertake a series of similar experiments with different GQDs, complemented by extensive characterization of the temperature- and magnetic field-dependent transport behavior. Furthermore, shot noise measurements might yield valuable insights into SYK dynamics \cite{Nikolaenko2023}. Our work demonstrates the possibility of disordered GQDs as an SYK platform and provides a first step toward experimental exploration of this novel quantum phase in solid-state systems.

\

\begin{acknowledgments}
The authors thank Bertrand Halperin and Alexander Altland for useful discussions. The major part of the experiment was supported by DOE (DE-SC0012260). L.E.A acknowledges support from ONR MURI (N00014-21-1-2537). K.W. and T.T. acknowledge support from the JSPS KAKENHI (Grant Numbers 20H00354 and 23H02052) and World Premier International Research Center Initiative (WPI), MEXT, Japan. L.S. and S.S. acknowledge support from U.S. National Science Foundation grant No. DMR-2245246. A.K. was supported by the Branco Weiss Society in Science, ETH Zurich, through the grant on flat bands, strong interactions and SYK physics, and Swiss National Science Foundation Grant No. CRSK-2\_221180. This work was performed, in part, at the Center for Nanoscale Systems (CNS), a member of the National Nanotechnology Infrastructure Network, which is supported by the NSF under Grant No. ECS-0335765. CNS is part of Harvard University.
\end{acknowledgments}

\bibliographystyle{apsrev4-1}

\bibliography{mainRefs}

\end{document}


\preprint{APS/123-QED}
    
\title{Supplemental Material for\\ ``Magneto-Thermoelectric Transport in\\ Graphene Quantum Dot with Strong Correlations''}

\author{Laurel E. Anderson}
\affiliation{Department of Physics, Harvard University, Cambridge, Massachusetts 02138, USA}
\author{Antti Laitinen}
\affiliation{Department of Physics, Harvard University, Cambridge, Massachusetts 02138, USA}
\author{Andrew Zimmerman}
\affiliation{Department of Physics, Harvard University, Cambridge, Massachusetts 02138, USA}
\author{Thomas Werkmeister}
\affiliation{Department of Applied Physics, Harvard University, Cambridge, Massachusetts 02138, USA}
\author{Leyna Shackleton}
\affiliation{Department of Physics, Harvard University, Cambridge, Massachusetts 02138, USA}
\author{Alexander Kruchkov}
\affiliation{Department of Physics, Harvard University, Cambridge, Massachusetts 02138, USA}
\affiliation{Department of Physics, Princeton University, Princeton, New Jersey 08544, USA}
\affiliation{Institute of Physics, \'Ecole Polytechnique F\'ed\'erale de Lausanne, Lausanne, CH 1015, Switzerland; and Branco Weiss Society in Science, ETH Zurich, Zurich, CH 8092, Switzerland}
\author{Takashi Taniguchi}
\affiliation{Research Center for Materials Nanoarchitectonics, National Institute for Materials Science,  1-1 Namiki, Tsukuba 305-0044, Japan}
\author{Kenji Watanabe}
\affiliation{Research Center for Electronic and Optical Materials, National Institute for Materials Science, 1-1 Namiki, Tsukuba 305-0044, Japan}
\author{Subir Sachdev}
\affiliation{Department of Physics, Harvard University, Cambridge, Massachusetts 02138, USA}
\author{Philip Kim}
\affiliation{Department of Physics, Harvard University, Cambridge, Massachusetts 02138, USA}
\affiliation{Department of Applied Physics, Harvard University, Cambridge, Massachusetts 02138, USA}
 \email{pkim@physics.harvard.edu}

\date{\today}
\raggedbottom
\setcounter{equation}{0}
\setcounter{figure}{0}
\setcounter{table}{0}
\setcounter{page}{1}
\makeatletter
\renewcommand{\theequation}{S\arabic{equation}}
\renewcommand{\thefigure}{S\arabic{figure}}
\renewcommand{\bibnumfmt}[1]{[S#1]}
\renewcommand{\citenumfont}[1]{S#1}

\newcommand*{\citen}{}
\DeclareRobustCommand*{\citen}[1]{%
  \begingroup
    \romannumeral-`\x 
    \setcitestyle{numbers}%
    \cite{#1}%
  \endgroup
}

\maketitle
\tableofcontents

\section{\label{section:fab}I: Device fabrication}
\begin{figure}
\centering
\includegraphics[width=0.8\textwidth]{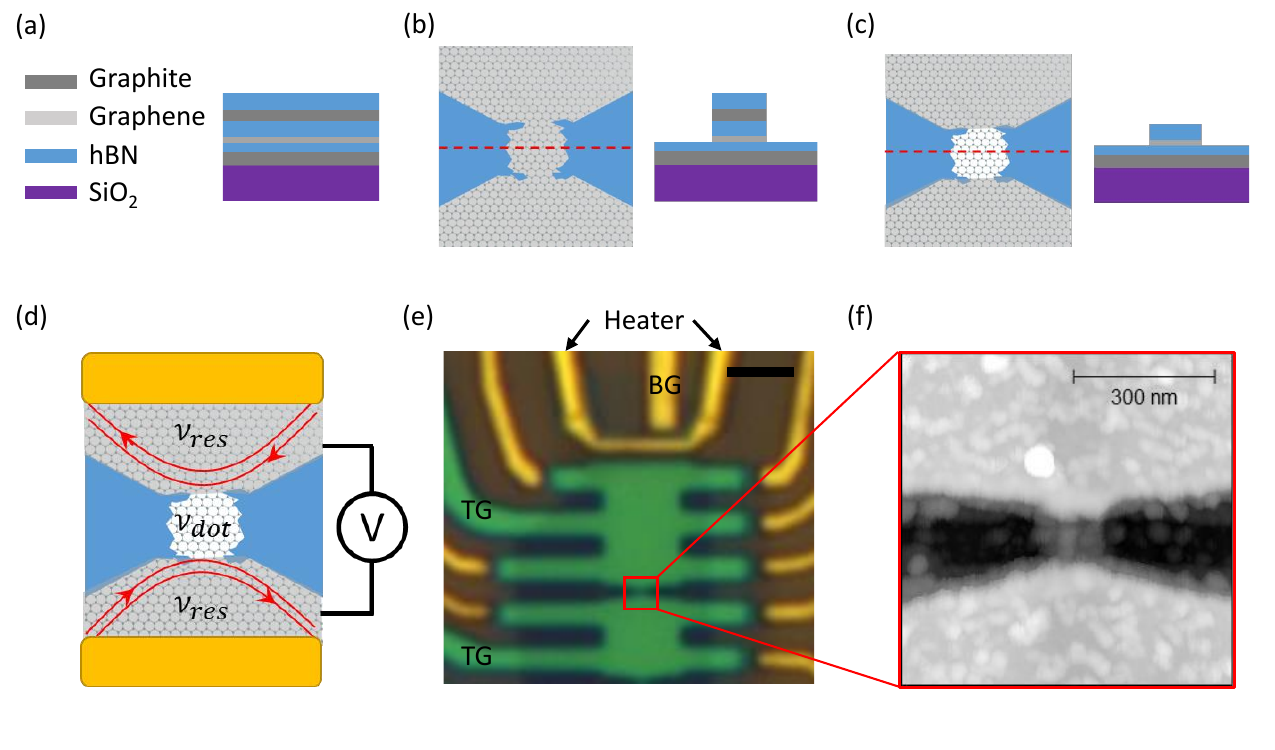}
\caption[Schematics and images of graphene quantum dot device fabrication and operation.]{\label{fig:dotfab} (a-c) Schematics of dot fabrication process. (a) Stack of van der Waals flakes used for the device: cover hBN, top graphite, top hBN, graphene, bottom hBN, bottom graphite (listed from top to bottom), resting on SiO$_2$/Si substrate. (b) Detail of dot definition by etching away cover hBN, top graphite, top hBN, and graphene, leaving an island of graphene connected to two larger graphene reservoirs. A cross-section of the etched stack along the red dashed line is shown at right. (c) Detail of definition of separate reservoir top gates and ungated dot region by etching cover hBN and top graphite above dot, with cross-section. (d) Simplified schematic of device operation: separate top gates independently tune the filling factors of the dot and reservoirs ($\nu_{dot}, \nu_{res}$), and measurement of the voltage across the dot upon application of electrical bias (temperature gradient) enables extraction of the electrical conductance (TEP). (e) Optical micrograph of an example device, with indications of substrate heater and bottom gate (BG) and top gate (TG) contacts. Scale bar is 2 $\mu$m. (f) Atomic force microscopy image of the red-outlined region in (e), showing etched GQD and separation of reservoir top gates.}
\end{figure}

The van der Waals heterostructure, comprising alternating layers of hexagonal boron nitride (hBN), graphite, hBN, graphene, hBN, and graphite was assembeled using the standard dry transfer technique~\cite{Wang2013}. Graphene, graphite and hBN are mechanically exfoliated using thermal release or Scotch tape onto doped Si substrates with a $285$ nm-thick SiO$_2$ layer on top. Thicknesses of all of the van der Waals flakes except for graphene (i.e. hBN and graphite) are confirmed using atomic force microscopy; monolayer graphene thickness is determined from optical contrast with the substrate~\cite{Cheng2019}. Typically, a large hBN flake is picked up with a PPC film on a PDMS stamp. This ``cover hBN'' flake is used to pick up subsequent layers: top graphite, top hBN, graphene, bottom hBN, and bottom graphite. The final stack is represented in Fig. \ref{fig:dotfab}(a). We use ``top hBN'' to denote the layer separating the graphene from top graphite to remain consistent with the nomenclature for ``bottom hBN'' separating graphene from bottom graphite. The PPC and stack were detached from the stamp by heating to $150$ $^{\circ}$C to melt the PPC, which is then removed by high-temperature vacuum annealing.

The shapes of the reservoirs, dot, and graphite gates, as well as the metal contacts and heater, were defined using electron beam lithography and reactive ion etching (RIE). A substrate heater, in the shape of a narrow ($\sim 400$ nm wide, few-micron long) rectangle of metal ($5$ nm Cr/$70$ nm Au) connected to two wider leads, is deposited on top of the stack in a region where the top graphite is etched away, with the rectangle running parallel to the shorter end of the Hall bar and roughly $500$ nm from the remaining top graphite gate. By passing a current through the heater, we can create a temperature gradient across the dot, provided that the narrow rectangle is the most resistive part of the heater and thus experiences the most Joule heating. The heterostructure is shaped into a Hall bar by etching with CHF$_3$ through a resist mask defined by e-beam lithography. A further lithography step defined the contact electrodes for the graphene and graphite gates, which were made by etching to expose a clean graphene/graphite edge, and then depositing a metallic trilayer ($2$ nm Cr/$8$ nm Pd/$50$ nm Au) using thermal evaporation (as described in Ref. \citen{Wang2013}).

The final steps are removing the top graphite from above the dot region and etching to define the dot. Chemically-selective RIE recipes also helps prevent accidentally etching through additional layers of the stack: SF$_6$ is used to etch hBN and weak ($30$ W) O$_2$ plasma for graphene and graphite. We first etch a $\sim 100$ nm-wide line across the width of the Hall bar through the cover hBN (using SF$_6$) and top graphite (using O$_2$). The top graphite etch is performed in $\sim 30$ second steps, stopping when the top graphite has been completely etched, as determined by resistance measurements between the two top graphite contacts between each step (Fig. \ref{fig:dotfab}(c)). A similar process is used to etch the dot, repeating the etch recipes and required etch times to remove the cover hBN and top graphite, followed by another SF$_6$ step to remove the top hBN and successive short ($15$ to $30$ second) weak O$_2$ steps to remove the graphene (Fig. \ref{fig:dotfab}(b)), stopping once the graphene two-terminal resistance across the dot region increased dramatically (typically from a few k$\Omega$ to tens of k$\Omega$). A final device with a $140$ nm wide by $90$ nm long dot (as determined by atomic force microscopy) is shown in Figure \ref{fig:dotfab}(e), with an atomic force microscopy image of the dot region in Figure \ref{fig:dotfab}(f). For some devices, multiple dots were defined along the same Hall bar, with independent contacts to the top graphite gates for each of the reservoir regions and typically a substrate heater on each end of the device, so that the heater closest to the dot being measured could be used to define the temperature gradient across the dot.

\section{\label{section:QH}II: Quantum Hall effect in reservoirs and dot at $B=10$ T}
\begin{figure}
    \centering
    \includegraphics[width=0.8\textwidth]{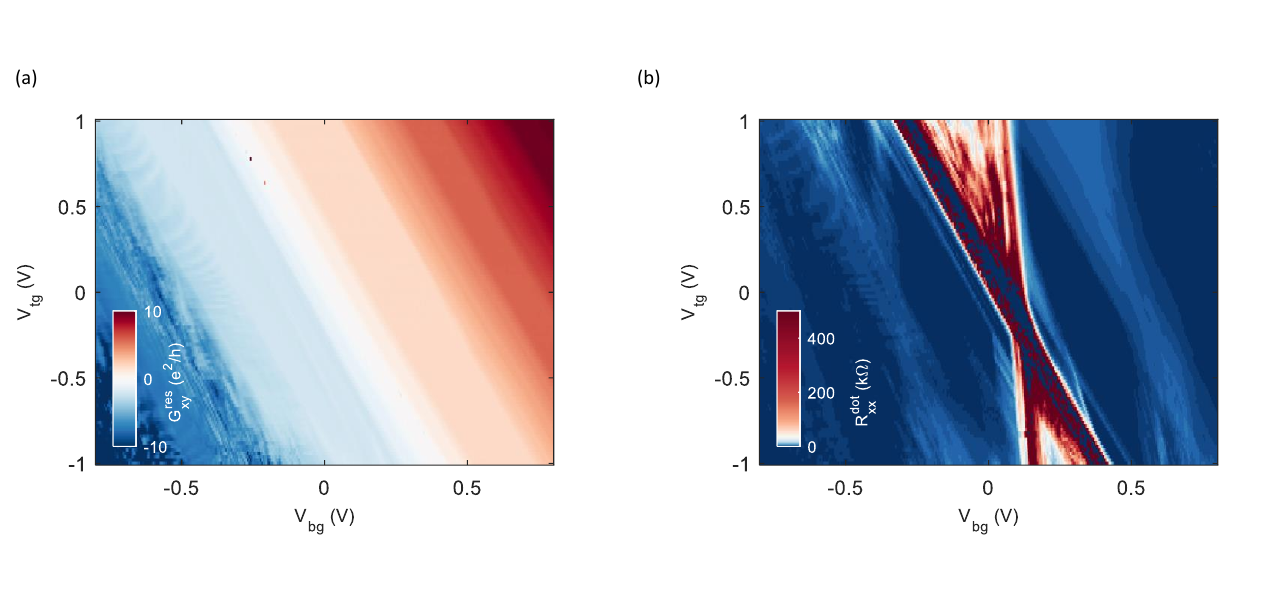}
    \caption{(a) Tranverse resistance of graphene reservoir $G_{xy}^{res}$ at $B=10$~T and $T_{bath}=350$~mK. (b) Simultaneously measured longitudinal resistance across GQD, $R_{xx}^{dot}$.}
    \label{fig:QHmaps10T}
\end{figure}
Investigating the device behavior in a strong perpendicular magnetic field allows us to understand how the different device regions interact and find the optimal gate voltage parameters for exploring the GQD contribution to transport effects. Figure \ref{fig:QHmaps10T}(a) shows the transverse resistance of one of the reservoirs, $G_{xy}^{res}=I/V_{xy}$. Robust quantum Hall plateaux with well-quantized conductance are evident at $\nu_{\rm res}=\pm$ 2, 6, 10, with smaller plateaux at other integer filling factors. Measuring electrical transport across the GQD (Fig. \ref{fig:QHmaps10T}(b)), we found that $R_{xx}^{dot}$ bears an initial resemblance to the zero magnetic field data (Fig. 1(b) of main text), with the high-resistance $npn$ regions becoming even more prominent and additional parallelogram-like features of near-zero resistance ($R_{xx}^{dot}\sim 30\:\Omega$) coinciding with parts of the most robust integer fillings in the reservoirs. The dot controls the transmission of quantum Hall edge states from one reservoir to another depending on its doping relative to the reservoirs, acting essentially as a quantum point contact\cite{Datta1995,Zimmermann2017}. In general, we focus on using a small number of quantum Hall edge states in the reservoirs to probe the behavior of the GQD, adjusting the bottom and top gate voltages simultaneously to keep the reservoirs at constant filling factor $\nu_{\rm res}=2$.

\section{\label{section:TEPmethod}III: TEP measurement technique}
In this section, we describe the measurement technique for the TEP experiments described in the main text. 
Unlike a measurement of resistance $R=V/I$, in which either voltage or current is directly applied by the experimenter and the other quantity is measured, determining the TEP $S=-\Delta V_{th}/\Delta T$ of a mesoscopic device requires careful measurement to extract both quantities. In our devices, running a current through a thin wire nearby causes Joule heating in the wire, which generates a temperature gradient across the sample. We must then measure both the thermally-induced voltage $\Delta V_{th}$ and the temperature gradient $\Delta T$. In this case, we used the AC ($2\omega$) technique \citen{GhahariKermani2014,Zuev2008} to detect $\Delta V_{th}$ and measurements of changes in the quantum Hall signal as a function of $T_{bath}$ and heater excitation to estimate the temperature gradient.
\subsection{A: AC TEP measurements}
The underlying principle of the AC method for TEP measurements is that applying an AC excitation to the heater generates a temperature gradient modulated at twice the excitation frequency, and this frequency dependence translates to $\Delta V_{th}$ as well. 
\begin{figure}
\centering
\includegraphics[width=0.8\textwidth]{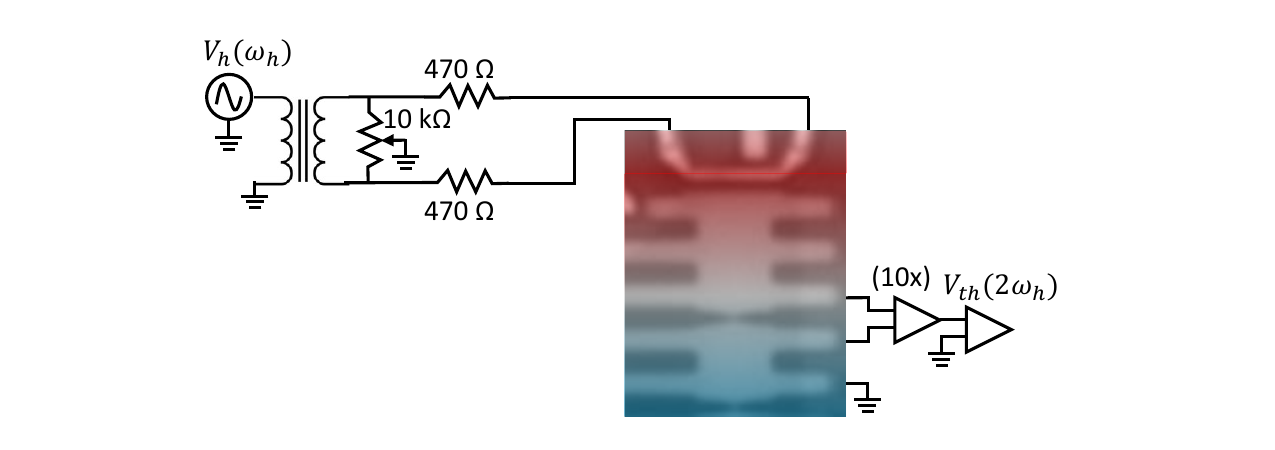}
\caption[Circuit schematic for dot TEP measurements.]{\label{fig:ACTEP} Circuit schematic for TEP measurements of the GQD.}
\end{figure}

Figure \ref{fig:ACTEP} shows the circuit used for AC TEP measurements across the dot. To avoid inadvertently creating a voltage offset between the heater and graphene channel, we symmetrically biased the heater using lock-in amplifier connected to a 1:1 transformer, with a tunable $10$ k$\Omega$ resistor between the output arms to tune the common mode potential. The output arms are connected to the ends of the heater through matched $470$ $\Omega$ resistors, to convert the voltage bias from the excitation lock-in to a current bias on the heater. The total two-terminal resistance of the heater side of the assembly, including the external resistors, line resistances, and heater resistance, was $R_h=1.453$ k$\Omega$, used to calculate $I_h$ when needed. $R_h$ was not found to change with temperature within the experimental range. The true amplitude $V'$ of the voltage excitation from the lock-in sine output is related to the root-mean-squared (rms) value on the lock-in display $V_{h}$ by
\begin{equation}
    V'=\sqrt{2}V_{h}\textrm{cos}(\omega_h t),
\end{equation}
where $\omega_h$ is the frequency of the AC excitation on the heater.

The temperature gradient $\Delta T$ resulting from the heater excitation is proportional to the Joule power,
\begin{equation}
    \Delta T=\zeta \frac{V'^2}{R_h}=2\frac{\zeta}{R_h}V_{h}^2\textrm{cos}^2(\omega_h t)= \frac{\zeta}{R_h}V_{h}^2\textrm{cos}(2 \omega_h t)+\frac{\zeta}{R_h}V_{h}^2,
\end{equation}
where $\zeta$ is a proportionality constant determined by the thermometry calibration measurements described in the next section. From the equation above, we see that applying an AC voltage to the heater at frequency $\omega_h$ generates a temperature gradient with an AC component at $2\omega_h$ with amplitude $\Delta T_{AC}=\frac{\zeta}{R_h}V_{h}^2$.

Once we have determined the corresponding $\Delta T$, we can find the TEP from voltage measurements at the second harmonic of the heater frequency:
\begin{equation}
    S_{AC}=\frac{\sqrt{2}\Delta V_{th}(2\omega_h)}{\Delta T},
\end{equation}
where the factor of $\sqrt{2}$ again comes from the fact that the lock-in amplifier detects the rms value of the voltage. The second harmonic signal is $\pi/2$ out of phase with the original excitation, and so should be measured on the Y channel of the lock-in amplifier. For our experiments, we typically also used a low-noise voltage preamplifier set to a gain of $10$ in order to improve the signal-to-noise ratio for measuring small thermal voltages, but we repeatedly confirmed that repeating the measurement without this preamplifier did not substantially affect the TEP reading.

\subsection{B: Temperature gradient estimation with quantum Hall thermometry}
\begin{figure}
\centering
\includegraphics[width=0.8\textwidth]{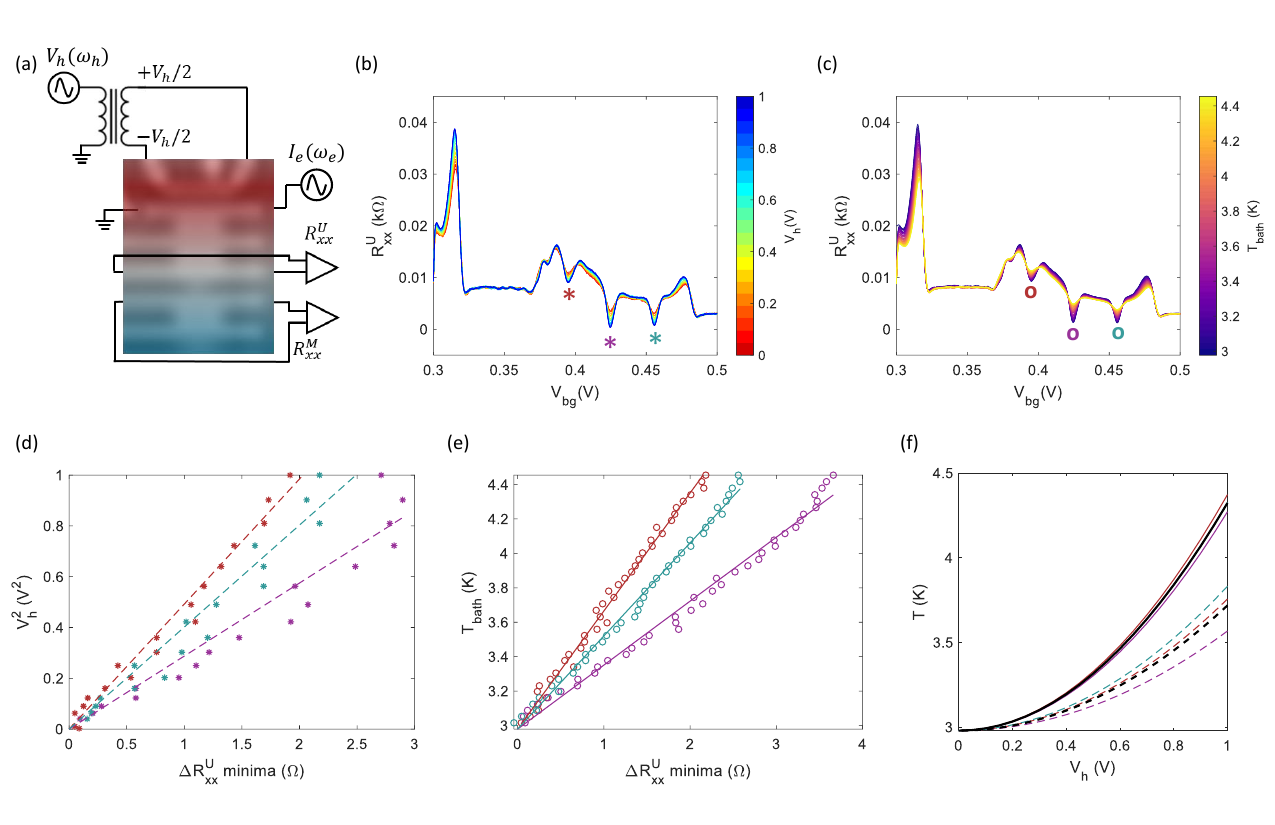}
\caption[Circuit schematic and example calibration data for temperature gradient estimation.]{\label{fig:Tcal}(a) Simplified circuit schematic for temperature gradient estimation: heater is symmetrically biased through a $1:1$ transformer, generating a temperature gradient across the sample proportional to $V_h^2$. Temperature increase across dot is estimated by measurement of changes in $R_{xx}$ on either side of the dot, using voltage probes at a constant distance from the heater, as a function of either $V_h$ or $T_{bath}$ (with no applied $V_h$). $R_{xx}$ is measured at frequency $\omega_e$. (b-e) Temperature calibration data, appearing as Fig. 3(c-d) in the main text. (b) Upper reservoir $R_{xx}$ ($R_{xx}^U$) as a function of $V_{bg}$ at various $V_h$, with $B=4$ T, $T_{bath}=3$ K and $V_{tg}=0$ V. $\nu_{res}=7,8,9$ minima indicated by red, purple and teal stars, respectively. (c) Same measurement as (b), but $V_h=0$ V and $T_{bath}$ is varied from $3$ K to $4.47$ K. $\nu_{res}=7,8,9$ minima indicated by circles. (d) $V_h^2$ versus change in $R_{xx}^U$ at $\nu_{res}=7,8,9$, with lines of best fit used for calibration. (e) $T_{bath}$ versus change in $R_{xx}^U$ at $\nu_{res}=7,8,9$, with lines of best fit. (f) Temperature increase from $T_{bath}$ in upper (solid lines) and middle (dashed lines) reservoirs as a function of $V_h$. Red, purple, and teal lines indicate individual estimates for $\nu_{res}=7,8,9$ minima, while bold black lines indicate the average of the estimates.}
\end{figure}
We now describe the method for determining the temperature gradient, using the calibration circuit shown in Fig. \ref{fig:Tcal}(a). The essential idea is to measure the longitudinal Hall resistance across two voltage probes that are at the same distance from the heater, and should therefore be at the same temperature. As the temperature increases, either globally due to a change in $T_{bath}$ or more locally due to an excitation in the heater generating a temperature gradient, the resistance values will change. In particular, the $R_{xx}$ minima at integer quantum Hall fillings should increase as
\begin{equation}
    R_{xx}^{min}\propto e^{-\Delta_a/k_B T},
\end{equation}
where $\Delta_a$ is the activation gap\cite{Giesbers2007}. For small changes in temperature $\delta T$, this can be approximated by $\Delta R_{xx}\propto \delta T$. We can therefore monitor changes in $R_{xx}$ as a function of either $V_h$ (at constant $T_{bath}$) or $T_{bath}$ (with no applied $V_h$). Since we have established that the temperature gradient due to the heater excitation is proportional to $V_h^2$, we will measure either $\Delta R_{xx}(\delta T_{bath})=A \delta T_{bath}$ or $\Delta R_{xx}(V_h)=B V_h^2$. Note that the coefficient $A$ contains the factor $\zeta/R_h$ defined in the previous section. By measuring simultaneously at several pairs of voltage probes, we can map the spatial temperature variations due to the heater excitation; in this case, we use the pairs of contacts in the upper and middle reservoirs closest to the dot ($R_{xx}^U$ and $R_{xx}^M$, respectively, in Fig. \ref{fig:Tcal}(a)). The lock-in excitation frequency for the $R_{xx}$ measurement $\omega_e=17.777$ Hz was chosen to avoid overlap with the heater frequency $\omega_h=23.333$ Hz or its second harmonic.

Some results of $R_{xx}$ measurements across the upper reservoir contacts at $B=4$ T as a function of $V_{bg}$ are shown in Fig. \ref{fig:Tcal}(b-c) for varying $V_h$ (b) and $T_{bath}$ (c), with an initial temperature of $T_{bath}=3$ K. We chose to perform the calibration at $B=4$ T due to the easy visibility of multiple $R_{xx}$ minima within a relatively small range of $V_{bg}$. We tracked the change in $R_{xx}$ at $\nu_{res}=7,\:8$, and $9$ in each reservoir as a function of $V_h^2$ and $T_{bath}$, shown in Fig. \ref{fig:Tcal}(d) and (e), respectively for the upper reservoir. Similar results were obtained for the middle reservoir, although the change in $R_{xx}^M$ as a function of $V_h^2$ was smaller due to the larger distance from the heater. By fitting the data, we could extract the calibration coefficients $A$ and $B$ for each $T_{bath}$, and thus find the conversion from applied $V_h$ to resulting temperature: $T=ABV_h^2$. The results for $T_{bath}=3$ K are shown in Fig. \ref{fig:Tcal}(f). We note that each $R_{xx}$ minimum gives a slightly different value for the calibration coefficients (indicated by corresponding colors in the figure), so the average value is used for final estimate of $\Delta T$, with any obvious outliers excluded.

At higher temperatures ($T_{bath}>6$ K), the $R_{xx}$ minima corresponding to symmetry-broken integer states had become nearly indistinguishable, so we used other features in the $R_{xx}$ data that varied linearly with respect to small changes in $T_{bath}$. While the use of other features rests on a slightly less robust theoretical foundation than $R_{xx}$ minima, we note this method has yielded reliable temperature estimates in other graphene quantum Hall systems\cite{Pierce2022}.

Several key assumptions underlie our use of this thermometry technique to convert the thermally-induced voltage $\Delta V_{th}(2\omega)$ to TEP. First, that the temperature gradient is imposed by the substrate and well-controlled by the heater, which has been consistently demonstrated in previous TEP measurements of graphene devices \cite{Checkelsky2009,Zuev2008,Ghahari2016}. Second, that the temperatures extracted from measurements at low magnetic field (typically $4$ T) can be used to calculate TEP at different magnetic fields from $0$ to $10$ T. This is sensible based on the first assumption, and borne out by the fact that measurements of the reservoirs with and without applied magnetic field conform to expectations based on theory and previous experiments (see Section \ref{section:reservoirS&Mott} below). Third, that the temperature gradient is not strongly affected by gate voltage. This is addressed in the discussion of Figure \ref{fig:Tcal}(d-f). We emphasize that each extracted temperature gradient was based on multiple quantum Hall features spanning a range of carrier densities in the reservoirs (and dot). Furthermore, the differences between calculated temperature gradients for different features are minimized by working at a small heater excitation $V_h$. This also has the benefit of minimizing $\Delta T/T$, which reduces the impact of any nonlinear effects on the TEP. 

Finally, we note the rescaling of calculated temperature gradients to account for the device geometry, which is particularly relevant for the dot. Since measurements of the dot TEP are all performed at $n_{res}$, the resistance (and TEP) of the reservoirs is constant, which means all the changes in the $\Delta V_{th}$ signal come from the dot. However, we calculate $\Delta T$ across the dot based on resistance measurements at the closest pairs of voltage probes in the upper and middle reservoirs, which are $d_{cal}=1.4\:\mu$m apart. This is much larger than the the ``length'' of the dot, $d=90$ nm (see Fig. \ref{fig:Mott0T}(b) below). The heater has no electrical connection to the graphene and should produce a nearly-linear temperature gradient across the region of interest\cite{Checkelsky2009,Zuev2008,Ghahari2016}, so the measured temperature difference $\Delta T$ must be rescaled. The same consideration does not apply to reservoir TEP measurements (see Figs. \ref{fig:Mott0T} and \ref{fig:RGS4Tres}), since $\Delta V_{th}$ and $\Delta T$ were measured using the same contacts and there is no constriction or junction between them to cause a more localized voltage drop. These approximations result in TEP results for both the dot and the reservoirs in good agreement with the Mott formula where it can be reasonably applied, as discussed below.

\section{\label{section:reservoirS&Mott}IV: Comparison of TEP with analytical expectations}

When TEP is only generated by electronic diffusion due to a temperature gradient, it can be semiclasically described by a relationship between the electrical conductivity and the TEP, known as the Mott formula\cite{Ashcroft1976,Jonson1984}:
\begin{equation}\label{eq:Mottbasic}
    S_{Mott}=-\frac{\pi^2}{3e}k_B^2T \left.\frac{1}{\sigma}\frac{d\sigma}{dE}\right|_{E_F},
\end{equation}
where $E_F$ is the Fermi energy. This relation can be rewritten in terms of experimentally-defined or measured quantities as 
\begin{equation}\label{eq:MottGrfull}
\begin{split}
      S_{Mott}=-\frac{\pi^2}{3e}k_B^2T\frac{1}{G}\frac{dG}{dV_{bg}}\left(\frac{dE_F}{dV_{bg}}\right)^{-1} \\
      =-\frac{\pi^2}{3e}k_B^2T\frac{1}{G}\frac{dG}{dV_{bg}}\frac{2}{\hbar v_F}\sqrt{\frac{e(V_{bg}-V_{bg}^0)}{\pi C_{dot}/A_{dot}}},  
\end{split}
\end{equation}
where $C_{dot}$ and $A_{dot}$ are the capacitance and area of the graphene dot, respectively.
Agreement between the Mott relation and measured TEP, in regimes where it is expected, can be a useful diagnostic of the robustness of the experimental protocols. However, the Mott formula relies on several assumptions, most particularly the ``relaxation time approximation,'' which is that the form of the non-equilibrium distribution function does not affect the distribution of electrons emerging from collisions or their collision rate~\cite{Ashcroft1976,GhahariKermani2014}. While this is generally true for elastic collisions (typical of electron-impurity scattering), electron-electron or electron-phonon interactions can produce inelastic collisions, which can cause violations of the Mott formula. The electron-phonon or ``phonon drag'' contribution dominates the TEP in GaAs above 0.6 K~\cite{Obloh1984,Fletcher1985,Ruf1988}. Graphene has relatively weak electron-phonon coupling \cite{Bolotin2009a,Efetov2010}, and this contribution has been found to be negligible in previous measurements \cite{Checkelsky2009,Zuev2008,Ghahari2016}. Electron-electron interactions play a more critical role in our system and are likely responsible, at least in part, for deviations from the Mott formula. 

\subsection{A: TEP agreement with Mott formula at zero magnetic field}

\begin{figure}
\centering
\includegraphics[width=0.9\textwidth]{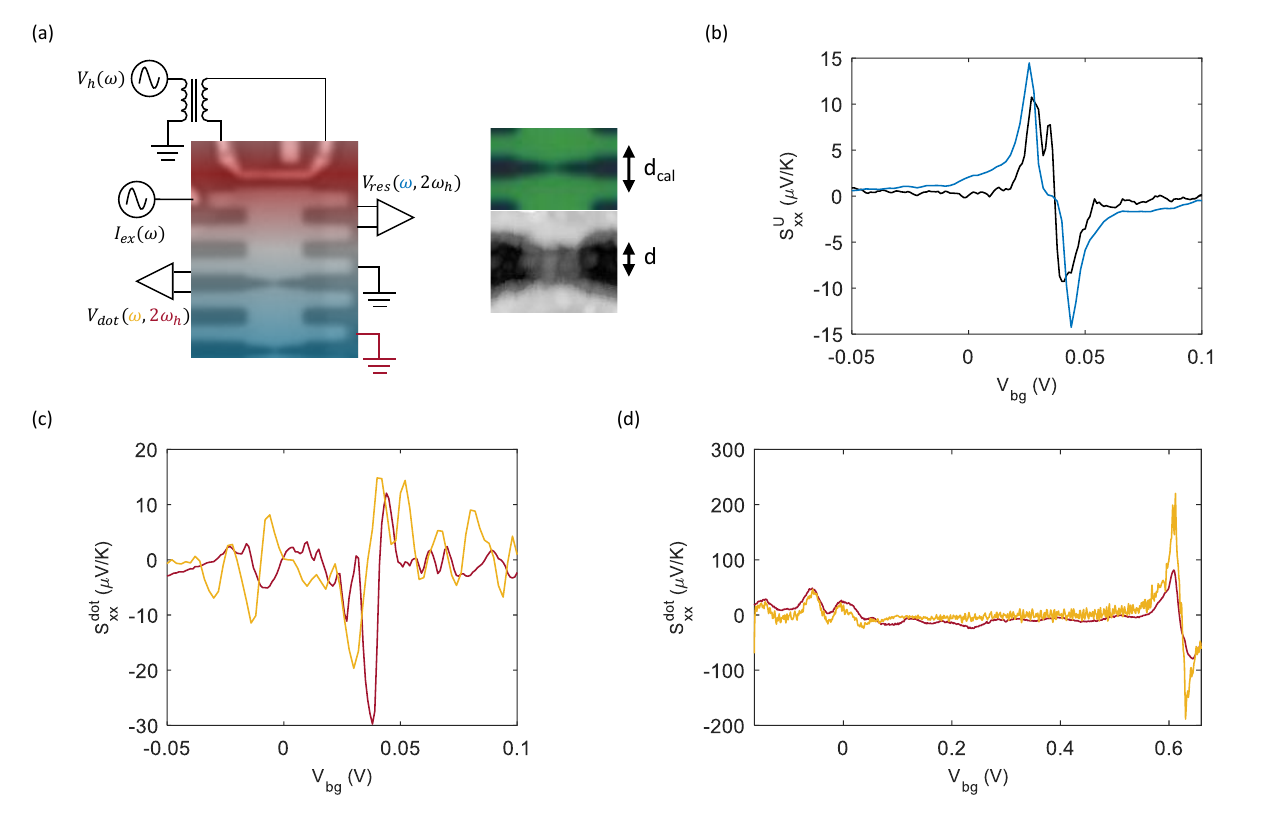}
\caption[Mott formula in reservoir and dot at B=0T]{\label{fig:Mott0T}(a): Configuration of TEP and resistance measurements of the reservoir (black and blue) and dot (red and yellow). Upper right: optical micrograph showing distance between calibration contacts $d_{cal}$ compared with atomic force micrograph showing actual dot length $d$. (b): TEP in upper reservoir $S_{xx}^U$ as a function of $V_{bg}$ at B = 0 T, $V_{tg}=0$ V, T = 3 K. Black: measured TEP $S_m$. Blue: calculation from Mott formula, $S_{xx}^{Mott}$ based on measurement of longitudinal resistance $R_{xx}^U$. (c): TEP across the dot $S_{xx}^{dot}$ as a function of $V_{bg}$ at B = 0 T, $V_{tg}=0$ V, T = 3 K. Red: measured TEP $S_m$. Yellow: calculation from Mott formula, $S_{xx}^{Mott}$. (d): $S_{xx}^{dot}$ as a function of $V_{bg}$ at constant reservoir density $n_{res}=4.8\times 10^{11}$ cm$^{-2}$ with B = 0 T, T = 31.4 K. Red: measured TEP $S_m$. Yellow: calculation from Mott formula, $S_{xx}^{Mott}$.}
\end{figure}

Figure \ref{fig:Mott0T}(a) shows the configurations for measurements of the longitudinal resistance (or equivalently conductance) and TEP in the upper reservoir, $R_{xx}^U$ and $S_{xx}^U$ respectively. The shape and magnitude of the TEP (Fig. \ref{fig:Mott0T}(b)) are consistent with previous measurements~\cite{GhahariKermani2014,Checkelsky2009,Ghahari2016}. We compare the measured TEP to the semiclassical Mott formula (eq. \ref{eq:MottGrfull}) and find good agreement between the measured TEP and the Mott formula prediction. Repeating the same TEP measurement across the dot at $T_{bath}=3$ K (Fig. \ref{fig:Mott0T}(c)), the quantitative level of agreement worsens, but many features and the overall magnitude remain consistent. The discrepancies may be due to complexities in the dot-reservoir coupling, as in this case the top gates were held at a fixed voltage during the measurement shown. We repeated the comparison at $T_{bath}=31.4$ K and constant reservoir density ($n_{res}=4.8\times 10^{11}$ cm$^{-2}$) and found near-perfect agreement (Fig. \ref{fig:Mott0T}(d)). These results suggest that our temperature calibration works reliably for the dot and the reservoirs in a wide range of temperatures.

\subsection{B: Quantum Hall TEP in reservoirs}
\begin{figure}
\centering
\includegraphics[width=0.9\textwidth]{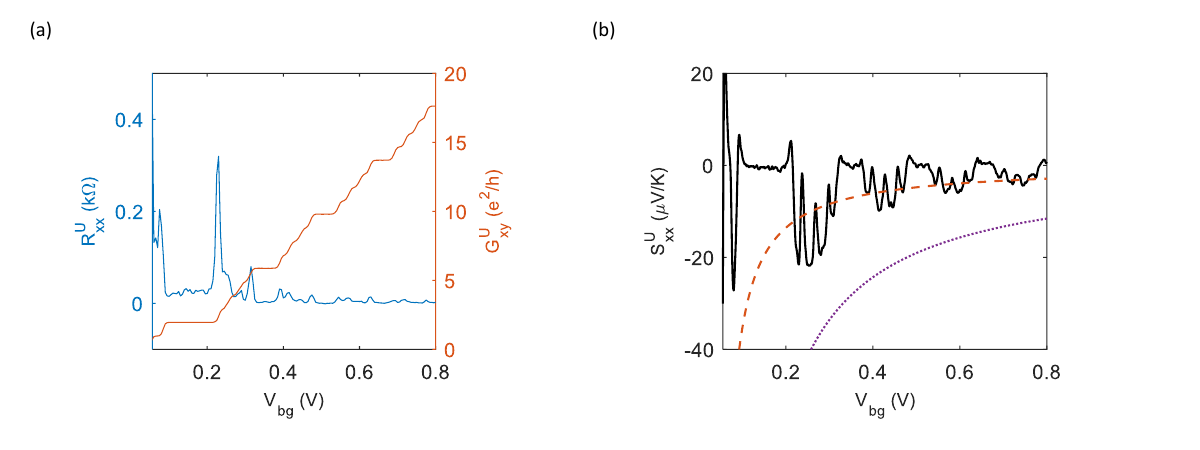}
\caption[Electronic and thermoelectronic transport in reservoir at B=4T]{\label{fig:RGS4Tres}(A): Quantum Hall electrical transport in upper reservoir as a function of $V_{bg}$ at B = 4 T, $V_{tg}=0$ V, T = 3 K. Blue: longitudinal resistance $R_{xx}^U$. Orange, right y-axis: transverse conductance $G_{xy}^U$ (b): TEP in upper reservoir $S_{xx}^{U}$ as a function of $V_{bg}$ at B = 4 T, $V_{tg}=0$ V, T = 3 K. Purple dotted and orange dashed lines correspond to equations \ref{eq:Smaxdegenerate} and \ref{eq:Smaxsplit}, respectively.}
\end{figure}

In the quantum Hall regime, single-particle, non-interacting theory~\cite{Jonson1984} for a disorder-free graphene system predicts peaks in the TEP at the center of each Landau level, with quantized maximum values given by
\begin{equation}\label{eq:Smaxdegenerate}
    S_{xx}^{max}=\frac{g_s k_B}{e}\frac{ln2}{\nu}
\end{equation}
where $g_s=4$ is the spin and valley degeneracy in graphene. With sufficiently high magnetic field (and low temperature) to break the degeneracy, this formula becomes
\begin{equation}\label{eq:Smaxsplit}
    S_{xx}^{max}=\frac{k_B}{e}\frac{ln2}{\nu}.
\end{equation}

Figure \ref{fig:RGS4Tres} shows the electronic ($R_{xx}$, $G_{xy}$) and thermoelectric ($S_{xx}$) quantum Hall response of a reservoir at $B=4$ T, with comparison to the theoretical expectations for $S_{xx}$ peaks with and without degeneracy breaking as given by equations \ref{eq:Smaxdegenerate} and 
\ref{eq:Smaxsplit}. The TEP peak values fall in between the two curves, suggesting incomplete lifting of the degeneracy at this relatively low magnetic field. Still, these results are in reasonable agreement with previous experiments~\cite{Zuev2008,GhahariKermani2014}.

\subsection{C: Dot TEP agreement with Mott formula with applied magnetic field}
\begin{figure}
\centering
\includegraphics[width=0.9\textwidth]{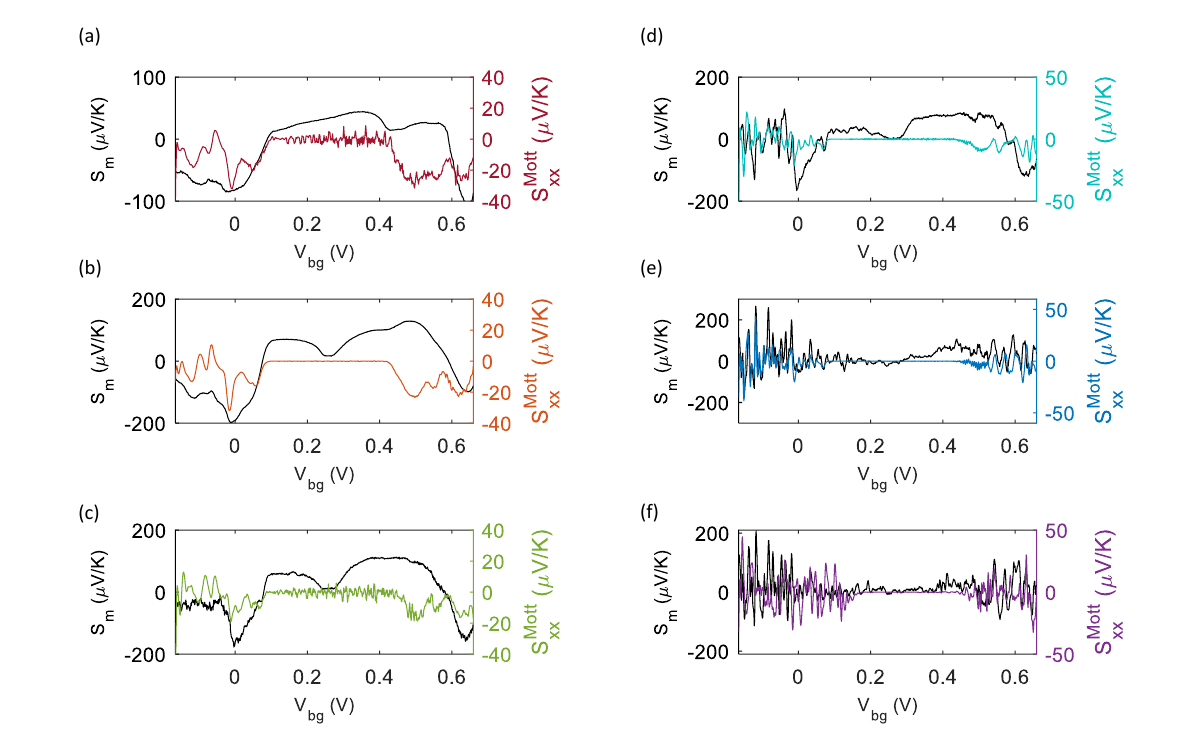}
\caption[Measured TEP and Mott formula calculation for dot at B=4T]{\label{fig:SMott10T_Tdep} Comparison of measured TEP $S_{m}$ (black lines) and Mott formula calculation $S_{xx}^{Mott}$ (colored lines) various temperatures with $B = 10$ T. Right y-axis is for Mott formula calculation. (a): 31.4 K (red). (b): 20 K (orange). (c): 10 K (green). (d): 5.25 K (turquoise). (e): 3 K (blue); reproduced from main text. (f): 1.41 K (purple).}
\end{figure}

\begin{figure}
\centering
\includegraphics[width=0.9\textwidth]{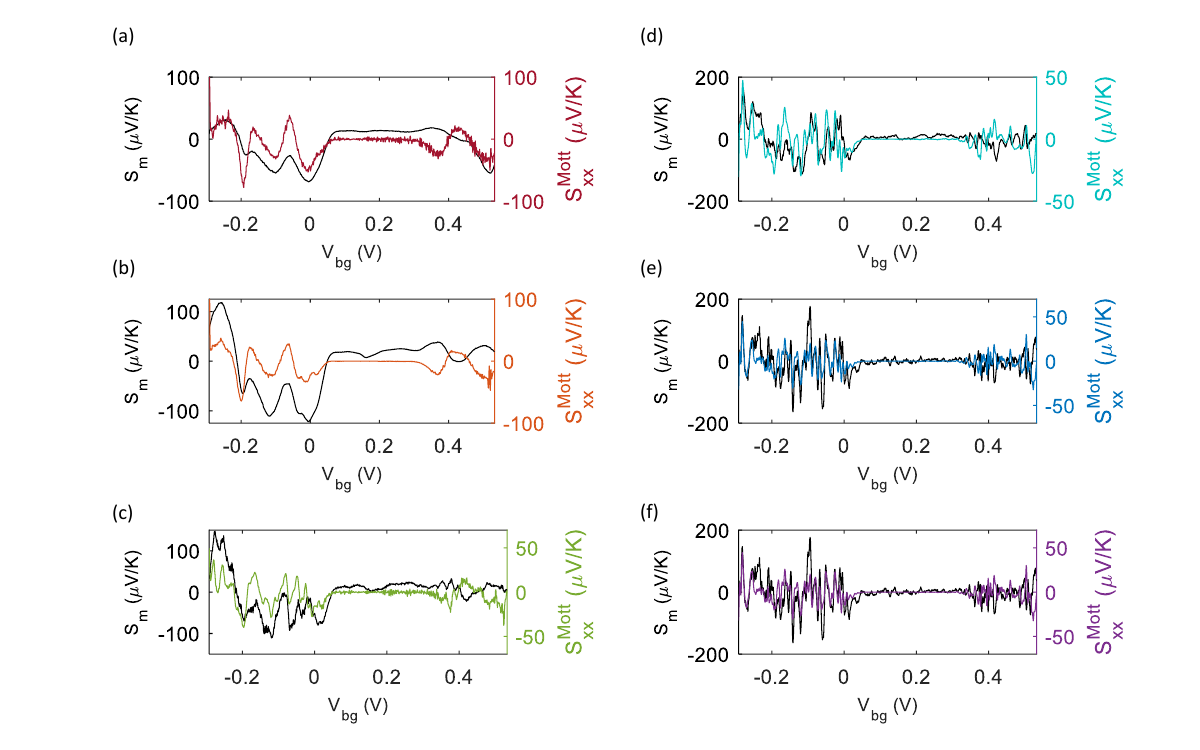}
\caption[Measured TEP and Mott formula calculation for dot at B=4T]{\label{fig:SMott4T_Tdep} Comparison of measured TEP $S_{m}$ (black lines) and Mott formula calculation $S_{xx}^{Mott}$ (colored lines) at various temperatures with $B = 4$ T. (a): 31.4 K (red). (b): 20 K (orange). (c): 10 K (green). (d): 5.25 K (turquoise). (e): 3 K (blue); reproduced from main text. (f): 1.41 K (purple)}
\end{figure}

We now consider an extended set of TEP data across the GQD at a range of temperatures with $B=10$ T and $B=4$ T. An applied magnetic field causes the response functions determining the TEP to become tensor quantities, leading to a generalized form of the Mott formula~\cite{Zuev2008,GhahariKermani2014,Ghahari2016}
\begin{equation}\label{eq:genMott}
    S_{ij}=-\frac{\pi^2}{3}\frac{k_B}{e}k_BT[\sigma^{-1}]_{il} \left[ \frac{\partial \sigma}{\partial \mu}\right]_{lj}.
\end{equation}
Comparisons between the measured $S_{xx}$ across the dot ($S_{m}$) and the Mott formula calculation based on $R_{xx}$ and $G_{xy}$ measurements measurements of the GQD ($S_{xx}^{Mott}$) are shown as a function of $V_{bg}$ in Figures~\ref{fig:SMott10T_Tdep} and ~\ref{fig:SMott4T_Tdep} for $B=10$ T and $B=4$ T, respectively. As before, $V_{tg}$ is simultaneously tuned to keep the reservoirs at $\nu_{res}=2$. A 3-point moving average is applied to the $R_{xx}$ data to reduce noise (particularly evident between $V_{bg}=0.1$ V and $0.5$ V, when the dot is more conductive), but the data points are spaced closely enough that this does not affect meaningful peak values or other features.

As discussed in the main text, the dot TEP at 10 T shows reasonable qualitative agreement between $S_{m}$ and $S_{xx}^{Mott}$ features for $V_{bg}<0$ (the $npn$ regime), some degree of correspondence for $0<V_{bg}<0.3$, and generally worse agreement at higher carrier density in the dot. On the quantitative level, $S_{m}$ exceeds the prediction of the Mott formula, with particularly large discrepancies at low temperatures. Although previous TEP measurements in graphene have generally found better agreement with the Mott relation at higher temperatures \cite{Zuev2011,Ghahari2016}, we actually observe larger deviations from the Mott formula prediction in the qualitative features. In particular, the measured TEP modulations in the $npn$ regime are much broader and smoother than would be expected from the derivative of the electrical conductance. As the temperature decreases below $\sim 5$ K (Fig.~\ref{fig:SMott10T_Tdep}(d-f)), the qualitative behavior of $S_{m}$ more closely resembles that of $S_{xx}^{Mott}$, with rapid fluctuations in the $npn$ and $nn'n$ regimes and approximately zero TEP when both reservoir QH edge states can pass through the dot.

At $B=4$ T, the agreement is better overall, although for $V_{bg}<-0.2$ there are more discrepancies in the relative peak magnitudes. Notably, the magnitudes of $S_{m}$ and $S_{xx}^{Mott}$ are much more similar at higher temperatures (Fig. \ref{fig:SMott4T_Tdep}(a-c)), with reasonably good matching of the qualitative features as well. The behavior at lower temperatures is similar to $B=10$ T, with $S_{mm}\gg S_{xx}^{Mott}$.

There are several factors that may contribute to the breakdown of the Mott formula in these devices in different regimes. In addition to the edge disorder effects discussed in the main text, quantum dots with an appreciable charging energy are expected to have a much larger TEP than would be expected from the Mott formula (by a factor of $k_BT/E_C$ in the ``classical'' regime $k_BT\gg E_C$)\cite{Beenakker1992}. This may account for some of the magnitude discrepancies at low temperatures. We also emphasize that the Mott formula is not expected to predict TEP in strongly-correlated systems, which may not obey the relaxation time approximation and may not be well-described by a quasiparticle picture. The more significant disagreement between the Mott formula and the measured TEP at higher temperatures at $B=10$ T compared to 4 T may indicate an increased importance of strongly-correlated effects.

\section{\label{section:4Tdata}V: Additional data at lower magnetic field}
\begin{figure}
\centering
\includegraphics[width=0.5\textwidth]{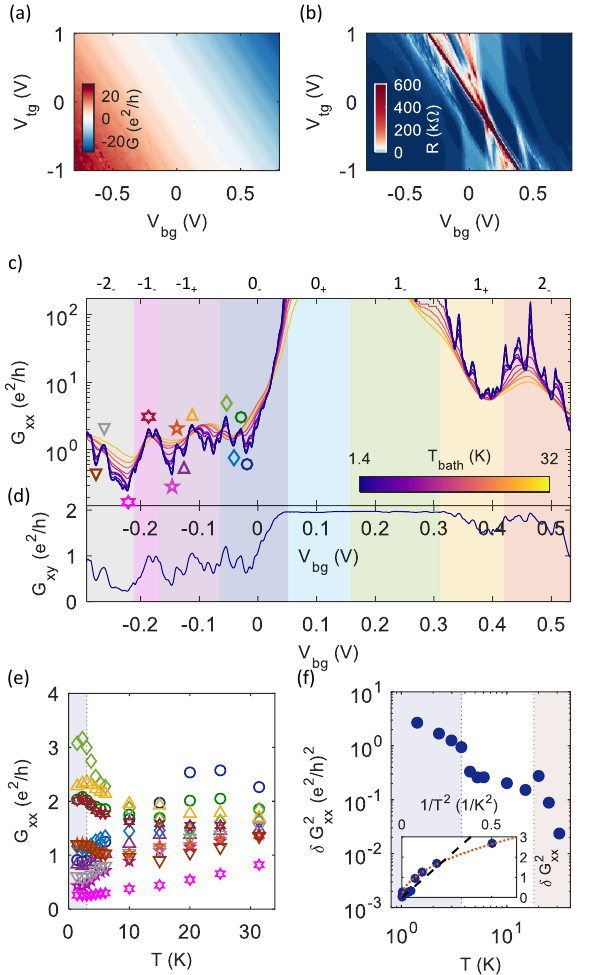}
\caption[Conductance data at B= 4T]{\label{fig:G4T}(a): Reservoir $G_{xy}$ at at $B=4$ T as a function of $V_{bg}$ and $V_{tg}$ at T=350 mK. (b): GQD $R_{xx}$ as a function of $V_{bg}$ and $V_{tg}$, measured simultaneously with (a). (c): $G_{xx}$ in the GQD at $B=4$ T with $V_{bg}$ and $V_{tg}$ simultaneously varied to maintain $\nu_{res}$=2, at a range of temperatures between 1.41 K and 31.4 K as indicated by the color scale. Shaded regions show the doping sectors for various Landau levels in the dot. Inset schematics illustrate the general behavior of the edge states in different doping sectors, as well as the voltages measured to determine $G_{xx}$ and $G_{xy}$. Open symbols mark minima plotted in (e). (d): $G_{xy}$ in the GQD measured along the same $V_{bg}$ and $V_{tg}$ values as (c) at T=1.41 K. (e): The five lowest minima and nearby peaks of $G_{dot}$ in the n=0 Landau level at $B=4$ T as a function of temperature. Blue dashed line marks onset of Fermi liquid behavior. (f): Variance of $G_{xx}$ in $0_-$ sector at $B=4$ T. Blue dashed line marks $T_1\sim 3.7$ K, while orange dashed line marks $T_2\sim18$ K. Inset: Variance of $G_{xx}$ versus $1/T^2$. Black dashed line and orange dotted line show $1/T^2$ and $1/T$ fits, respectively. $1/T^2$ fit is for $T>3$ K.}
\end{figure}
\begin{figure}
\centering
\includegraphics[width=0.5\textwidth]{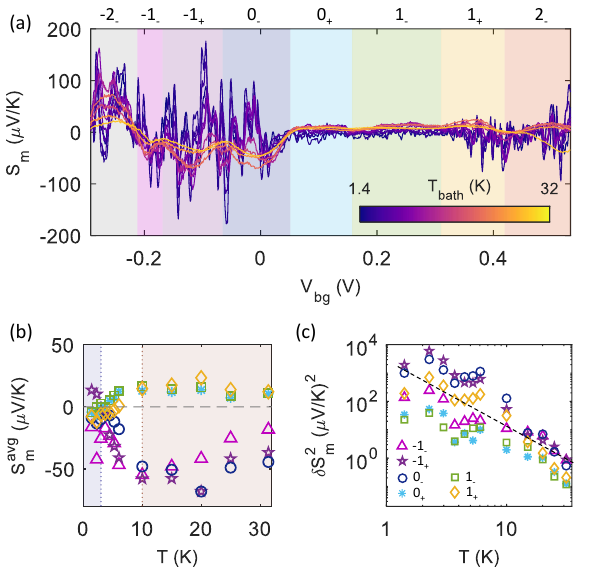}
\caption[TEP data at B=4T]{\label{fig:Sdot4T}(a): $S_{m}$ at $B=4$ T with $\nu_{res}$=2 at a range of temperatures between 1.41 K and 31.4 K. Shading indicates doping sectors for various Landau levels in the dot. (b): Average value of $S_{m}$ as a function of bath temperature in the sectors highlighted by colors in (a). (c): Variance of $S_{m}$ as a function of bath temperature for the sectors highlighted in (a). Dashed line shows $1/T^2$ as a guide to the eye.}
\end{figure}

To provide additional context for the results presented in Figures 2 and 4 of the main text, we repeated the same analysis for measurements of the same device at $B=4$ T, shown in Figures \ref{fig:G4T} and \ref{fig:Sdot4T}. Qualitatively, we observe much of the same behavior at this lower magnetic field, with some changes in the associated temperature scales. At 4 T, the typical sequence of IQH states ($\nu=2,6,10...$) remains strong in the reservoirs (Fig.~\ref{fig:G4T}(a)), and the same pattern of high-resistance $npn$/$pnp$ regions, low-resistance regions where $\nu_{dot}=\nu_{res}$, and re-emergent resistance in $nn'n$ regions occurs for GQD $R_{xx}$ as a function of the top and bottom gate voltages (Fig.~\ref{fig:G4T}(b)). A larger range of $\nu_{dot}$ and $\nu_{res}$ is accessible in the same gate voltage range, due to the lower magnetic field. 

These trends are clear in line cuts of $G_{xx}$ and $G_{xy}$ in the GQD (Fig.~\ref{fig:G4T}(c-d)). The IQH states are generally less robust to increases in temperature than at higher magnetic field (as expected from the reduced Landau level separation), which may account for the reduced conductance on the higher-dot-doping side of $\nu_{dot}=2$ ($1_-$ sector). However, we note that the $\nu_{res}=2$ state remains well-developed even at the highest temperatures tested. Tracking the temperature dependence of local conductance extrema in the $npn$ regime (Fig.~\ref{fig:G4T}(e)), we see the minima lifting faster after $T\sim 3$ K, and maxima lowering at the same temperature after initially rising or remaining relatively constant. The values do not collapse onto a single line as they do at $B=10$ T, but this is likely because we are tracking a larger range of filling factors in the dot, and thus sampling more combinations of junction-forming IQH states in the dot and reservoirs \cite{Abanin2007,Ozyilmaz2007}.

In the conductance variance (Fig.~\ref{fig:G4T}(f)), we see power law scaling $\sim 1/T$ for $T\leq 20$ K, then a much steeper temperature dependence starting at 20 K. There are significant jumps at $3.7$ K $< T < 4.2$ K and $15$ K $< T < 20$ K. Plotting the variance versus $1/T^2$ (inset of Fig.~\ref{fig:G4T}(f)), we find good agreement with $1/T$ dependence at low $T$, although this clearly breaks down at higher temperatures. This stands in contrast to the behavior at $B=10$ T, where we see a regime of $1/T^2$ dependence for $3$ K $< T < 10$ K and near-constant conductance variance elsewhere. This difference suggests stronger coupling between the dot and reservoirs at lower magnetic field \cite{Shackleton2023}, which is reasonable given the smaller Landau level spacing and the evidence of relatively strong dot-reservoir coupling without magnetic field. 

The behavior of the dot TEP at $B=4$ T with varying dot carrier density and temperature is relatively similar to our observations at higher magnetic field reported in the main text. We note that the anomalous sign of the TEP at low carrier densities in the dot persists beyond the lowest Landau level to $n=\pm 1$. 
Within each Landau level sector, we see the average TEP values (Fig.~\ref{fig:Sdot4T}(b)) converge to near zero at the transition temperature $T_1 = 3$ K, increase until $T_2 = 10$ K, and then decrease slowly as temperature increases further. As at $B=10$ T, the TEP variances in each sector (Fig. \ref{fig:Sdot4T}(c)) decrease with temperature across the entire measured range. Below 10 K, there is a greater spread in variance values, with the overall trend consistent with scaling between $1/T$ and $1/T^2$. For $T\geq 10$ K, the values fall more closely together, following $\sim 1/T^2$. We again note that a relatively small estimated population of SYK modes in the dot (approximately 13 at $B=4$ T) may result in paradigmatic SYK behavior being masked by finite-size effects~\cite{Shackleton2023}.

\section{\label{section:dotB}VI: Additional data from second GQD}
\begin{figure}
\centering
\includegraphics[width=0.8\textwidth]{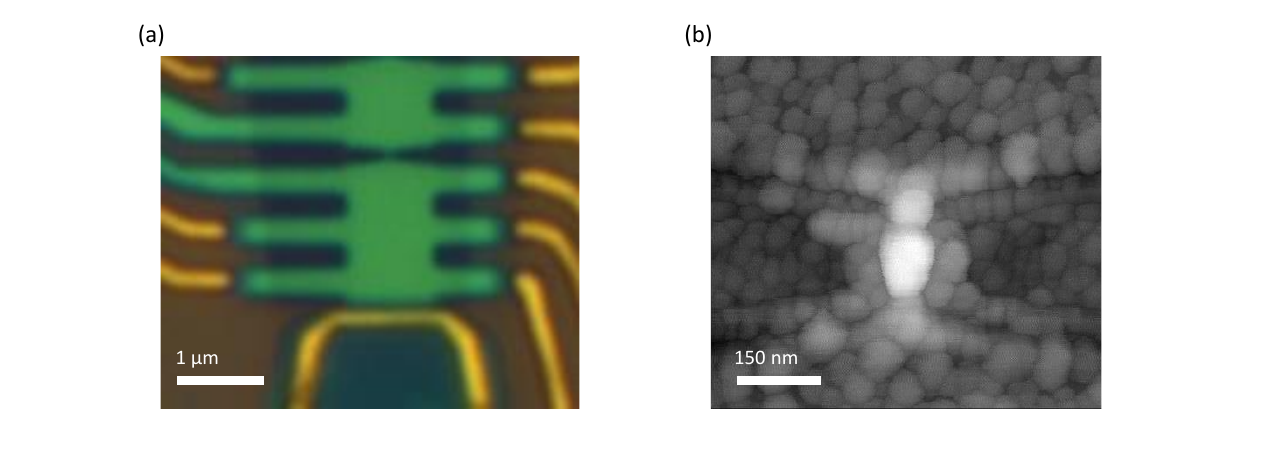}
\caption[Optical and AFM images of second device.]{\label{fig:dotBpics}(a): Optical micrograph of second GQD and reservoirs with heater. Extended heterostructure leads on left side are contacts to the reservoir top gates. (b): Atomic force micrograph of second GQD.}
\end{figure}
We repeated the same measurement protocol for electronic and thermoelectric transport measurements on a second dot. It was etched from a different section of the same heterostructure, with a separate heater fabricated nearby to facilitate efficient generation of a temperature gradient (Fig.~\ref{fig:dotBpics}(a)). The order of reactive ion etching steps for shaping the dot and removing the top graphite gate above it was also reversed, resulting in additional residue buildup, particularly above the GQD (Fig.~\ref{fig:dotBpics}(b)). However, the reservoir top gates were confirmed to be electrically disconnected and separated on both sides of the GQD by atomic force microscopy.

\begin{figure}
\centering
\includegraphics[width=0.5\textwidth]{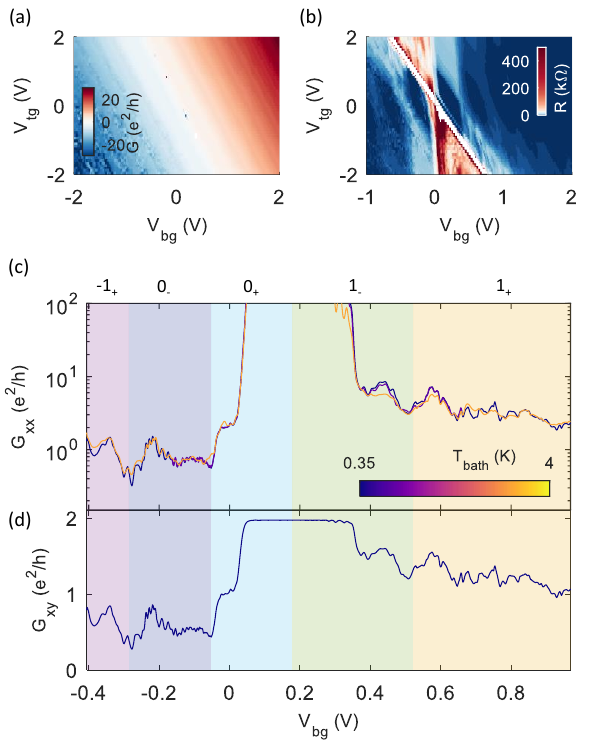}
\caption[Conductance data in second device]{\label{fig:G10TB}(a): Reservoir $G_{xy}$ in second device at $B=10$ T as a function of $V_{bg}$ and $V_{tg}$ at $T=350$ mK. (b): $R_{xx}$ in the GQD as a function of $V_{bg}$ and $V_{tg}$, measured simultaneously with (a). (c): $G_{xx}$ in the GQD at $B=10$ T with $V_{bg}$ and $V_{tg}$ simultaneously varied to maintain $\nu_{res}=2$, at a range of temperatures between 350 mK and 3.7 K as indicated by the color scale. (d): $G_{xy}$ measured along the same $V_{bg}$ and $V_{tg}$ values as (c) at $T=350$ mK.}
\end{figure}

Electrical conductance measurements of the second dot at $B=10$ T are shown in Figure~\ref{fig:G10TB}. The gate dielectric breakdown voltages were higher in this set of measurements, enabling us to probe an expanded range of filling factors in both the dot and reservoirs (Fig.~\ref{fig:G10TB}(a-b)). Similar patterns of gate voltage dependence are evident in this device relative to the first, although $R_{xx}$ in the GQD is generally lower in the $npn$ and $pnp$ regimes. The bottom hBN gate dielectric suffered a catastrophic breakdown during the temperature dependence measurements, limiting the available data set to $T<3.7$ K for the conductance and $T<4.5$ K for the TEP. 
The relationship between the electrical conductance and dot carrier density shows the same patterns in the $npn$, full-IQH-transmission, and $nn'n$ regimes (Fig.~\ref{fig:G10TB}(c-d)). Unlike the first dot, there is a clear conductance plateau at $\nu_{dot}=1$, suggesting this dot may be less dominated by localized charging in the low temperature regime. Unfortunately, the data set acquired before breakdown was not sufficient to meaningfully track the evolution of the conductance extrema or the variance with temperature.

\begin{figure}
\centering
\includegraphics[width=0.5\textwidth]{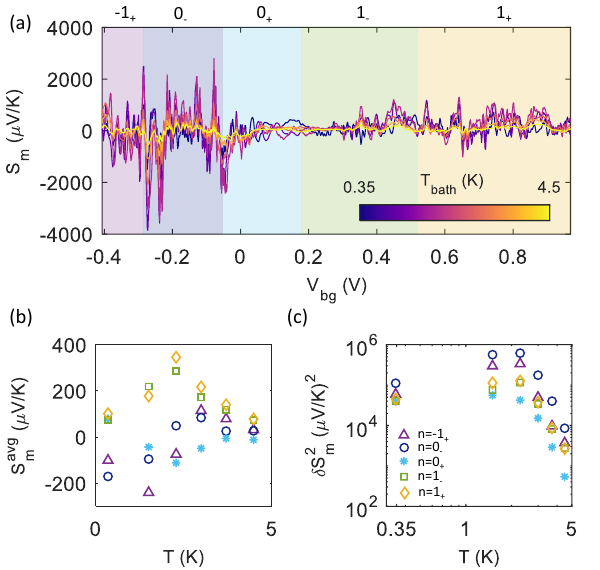}
\caption[TEP data in second device]{\label{fig:S10TB}(a): $S_{m}$ in second device at $B=10$ T with $\nu_{res}=2$ at a range of temperatures between 350 mK and 4.47 K. Shading indicates doping sectors for various Landau levels in the dot. (b): Average value of $S_{m}$ as a function of bath temperature in the sectors highlighted by colors in (a). (c): Variance of $S_{m}$ as a function of bath temperature for the sectors highlighted in (a).}
\end{figure}

Figure~\ref{fig:S10TB} shows TEP measured at $B=10$ T at temperatures between 350 mK and 4.5 K. The temperature gradient $\Delta T$ across the dot was estimated using the method described in Section III. Calibration measurements were performed using pairs of reservoir contacts at greater distances from the GQD compared to the first device because of the geometry of the top gate electrodes (see Fig.~\ref{fig:dotBpics}(a)). This leads to a larger geometric correction to $\Delta T$, and a greater possible effect of non-uniformity in the temperature profile (for example, if the temperature gradient is larger across the dot than across the reservoirs). Since this method likely underestimates $\Delta T$ across the dot, the reported TEP should be regarded as an upper bound. 

As with the conductance measurements, we see qualitatively similar dependence of the dot TEP on magnetic field and temperature between the first and second devices. The near-complete transmission of reservoir IQH states through the dot in the $0_+$ and $1_-$ Landau level sectors generates very little TEP, and elsewhere the rapid fluctuations at the lowest temperatures begin to give way to more gradual modulations by $4.5$ K. The average $S_{m}$ in each sector (Fig.~\ref{fig:S10TB}(b)) shows a similar tendency toward more extreme values with increasing temperature up to (typically) $\sim 2.5$ K, then all trend toward zero as temperature increases further. Interestingly, the $0_-$ sector show an increasing value with a change in sign from negative to positive up to 3 K, before reducing toward zero. This may be a side effect of larger variance in this sector than the others (Fig.~\ref{fig:S10TB}(c)). Above 1.4 K, all sectors show similar scaling of $\delta^2 S_m$ to the first device; however, the variance seems to flatten or even decrease as temperature is lowered below that. While this may reflect a separate regime of behavior emerging at very low temperatures, the TEP data at 350 mK was also obtained with a larger thermal ``bias'' $\Delta T/T\approx 14\%$, compared to $\Delta T/T\sim 5\%$ that was achievable for higher bath temperatures. Although still a small excitation, it is possible that this relatively larger thermal gradient may artificially reduce the TEP variance. Without higher-temperature results, it is difficult to draw any significant conclusions about the potential existence of SYK dynamics in this device. However, we can at least demonstrate that the low-temperature behavior remains quite consistent between devices.

\clearpage
\bibliographystyle{apsrev4-2}
\bibliography{supplementalRefs}